\documentclass{elsarticle}

\usepackage[utf8]{inputenc}
\usepackage{lscape}
\usepackage{amssymb}
\usepackage{hyperref}
\usepackage{listings}
\usepackage{pifont}
\usepackage{subcaption}
\usepackage{calc}
\usepackage{url}
\usepackage{amstext}
\usepackage{multicol}
\usepackage{graphicx}
\usepackage{booktabs}
\usepackage{multirow}
\usepackage{bbding}
\usepackage{arydshln}
\usepackage{soul}

\usepackage{flushend}
\usepackage{ifthen}
\usepackage{tcolorbox}
\usepackage{paralist}

\journal{Journal of Systems and Software}

\begin{document}

\begin{frontmatter}

\title{Assessing the Linguistic Quality of REST APIs for IoT Applications}

\author[1]{Francis Palma\corref{cor1}}
\ead{francis.palma@lnu.se}
\author[1]{Tobias Olsson}
\ead{tobias.olsson@lnu.se}
\author[1]{Anna Wingkvist}
\ead{anna.wingkvist@lnu.se}
\author[2]{Javier Gonzalez-Huerta}
\ead{javier.gonzalez.huerta@bth.se}
\cortext[cor1]{Corresponding author}

\address[1]{Department of Computer Science and Media Technology, Linnaeus University}
\address[2]{Department of Software Engineering, Blekinge Institute of Technology, Sweden}

\begin{abstract}
Internet of Things (IoT) is a growing technology that relies on connected `things' that gather data from peer devices and send data to servers via APIs (Application Programming Interfaces). The design quality of those APIs has a direct impact on their understandability and reusability.
This study focuses on the linguistic design quality of REST APIs for IoT applications and assesses their linguistic quality by performing the detection of linguistic patterns and antipatterns in REST APIs for IoT applications. \textit{Linguistic antipatterns} are considered poor practices in the naming, documentation, and choice of identifiers. In contrast, \textit{linguistic patterns} represent best practices to APIs design. The linguistic patterns and their corresponding antipatterns are hence contrasting pairs.
We propose the SARAv2 (Semantic Analysis of REST APIs version two) approach to perform syntactic and semantic analyses of REST APIs for IoT applications. Based on the SARAv2 approach, we develop the \textsf{REST-Ling} tool and empirically validate the detection results of nine linguistic antipatterns. We analyse 19 REST APIs for IoT applications.
Our detection results show that the linguistic antipatterns are prevalent and the \textsf{REST-Ling} tool can detect linguistic patterns and antipatterns in REST APIs for IoT applications with an average accuracy of over 80\%. Moreover, the tool performs the detection of linguistic antipatterns on average in the order of seconds, \textit{i.e.}, 8.396 seconds.
We found that APIs generally follow good linguistic practices, although the prevalence of poor practices exists. 
\end{abstract}


\end{frontmatter}


\section{Introduction}\label{sec:intro}

The Internet of Things (IoT) is an emerging technology that relies on networks of `things' - smart computing devices - communicating with each other over the Internet. These `connected things' (\textit{i.e.}, IoT devices) frequently gather data and send them to servers or receive data from peer devices and act on them~\cite{Madakam2015}.

REpresentational State Transfer (REST)~\cite{Fielding2000} is the \textit{de facto} standard to design, develop, and deploy IoT-based applications in cloud environments. The application programming interfaces (APIs) for IoT applications are designed following the REST principles by the IoT vendors. The Message Queuing Telemetry Transport (MQTT) protocol is also supported by some APIs and the topic structure of MQTT could be seen as an alternative to the API design in REST. However, the offered functionality is often more limited in its scope, and it is often only recommended to use MQTT for certain applications where the lightweight nature of the protocol is necessary. In this article, for the sake of simplicity, from this point on, we refer to the \texttt{`REST APIs for IoT applications'} as \texttt{`IoT APIs'}. Client developers design and develop applications for IoT devices using vendor-provided IoT APIs that interact/communicate with peers and gateway servers. The design quality of IoT APIs has a direct impact on their understandability and reusability. Well-designed and named APIs may attract client developers more than poorly designed or named APIs~\cite{Masse2012} because they must understand the providers' APIs while integrating their services.

In previous works, we have performed analyses on the APIs for Web applications (\textit{e.g.}, Facebook, YouTube, Dropbox, etc.) and Cloud services (\textit{e.g.}, Open Stack) \cite{PalmaICSOC20014, Palma2015AreRA, Petrillo2016}. Yet, no such study has been performed to investigate how well the APIs dedicated to the IoT applications are designed in terms of linguistic quality. To measure the linguistic quality, we perform syntactic and semantic analysis of the URIs and their documentation. According to Wilhelm  et al.~\cite{wilhelm2013compiler}, in the context of computer programs, syntactic analysis recognises the syntactic structure of the programs. In contrast, semantic analysis helps determine properties and check conditions relevant to the programs' well-formedness according to the programming language rules. Thus, in our study, the syntactic analysis concerns the syntactic structure of the resource URIs and the semantic analysis checks for the well-formedness of the resource URIs according to the good API design practices defined in the literature ~\cite{Masse2012,Petrillo2016,Petrillo10.1007978-3-319-94959-816,alliance2012guidelines,10.1007/978-3-319-94959-816, 7930199, Hausenblas2011,Brabra10.1007978}. We perform a study on APIs solely designed for IoT devices and applications.

The linguistic and semantic relations among the `things', services, and parameters are as crucial in IoT APIs as in APIs for Web applications~\cite{RESTfulBestPractices}. The lack of such relations and/or poor naming may degrade the overall design of IoT APIs and translate into \textit{linguistic antipatterns}. In the context of IoT APIs, linguistic antipatterns are poor solutions to common URI (Uniform Resource Identifier) design problems, which may hinder the consumption and reuse of IoT APIs by client developers; and the maintenance and evolution of IoT APIs by API vendors. Conversely, linguistic patterns represent good solutions to common URI design problems and facilitate the consumption and maintenance of IoT APIs. Thus, the linguistic patterns and their corresponding antipatterns are contrasting pairs.

An example of a poor practice is \textcolor{red}{\begin{scriptsize}\ding{54}\end{scriptsize}}\textit{Inconsistent Documentation}\footnote{We use the \textcolor{red}{\ding{54}} symbol to refer an \textit{antipattern}} where a resource URI (together with the HTTP method) is in contradiction with its documentation. In the IBM Watson IoT, the POST method with \url{/bulk/devices/remove} URI is in contradiction with its documentation\footnote{ \texttt{Delete multiple devices. Delete multiple devices, each request can contain a maximum of 512 kB}}. In REST, the POST method should be used to create something. The presence of \textit{Inconsistent Documentation} may confuse IoT client developers who require clear and uniform resource specifications. The understandability and usability of the API might be hindered if this linguistic antipattern exists. In contrast,  \textcolor{teal}{\begin{scriptsize}\ding{52}\end{scriptsize}}\textit{Consistent Documentation}\footnote{We use the \textcolor{teal}{\ding{52}} symbol to refer a \textit{pattern}} is a linguistic pattern where a URI is in line with its documentation. The URI \url{/draft/physicalinterfaces/{physicalInterfaceId}} with the HTTP DELETE method from the IBM Watson IoT API is an example of this pattern with its documentation\footnote{\texttt{Delete a draft physical interface. Deletes the draft physical interface with the specified id from the draft physical interface}}.

In this research, we propose the SARAv2 approach (Semantic Analysis of REST APIs version two) as an extension to our previous approach, SARA~\cite{Palma2017SemanticAO}. SARAv2 is can perform semantic analysis of REST APIs in general, and therefore also IoT REST APIs, aiming to assess their linguistic quality, by detecting linguistic patterns and antipatterns. Being inspired from the object-oriented domain~\cite{Arnaoudova2013,Arnaoudova2016}, we define three new linguistic patterns and their corresponding linguistic antipatterns, namely \textit{Consistent \textit{vs.} Inconsistent Documentation}, \textit{Versioned \textit{vs.} Unversioned URIs}, and \textit{Standard vs. Non-standard URI}.

We develop the \textsf{REST-Ling} tool as the implementation of the SARAv2 approach. \textsf{REST-Ling} is a web application that automates the detection of linguistic patterns and antipatterns. Applying the \textsf{REST-Ling} tool, we perform the detection of nine linguistic patterns and their corresponding antipatterns in 1,102 URIs from 19 IoT APIs, \textit{e.g.}, Amazon, Cisco, Google, IBM, Microsoft, Samsung. \textsf{REST-Ling} utilises various NLP techniques including the traditional WordNet \cite{wordnet} and Stanford's CoreNLP ~\cite{manning-EtAl:2014:P14-5} general-purpose English dictionaries analysed with Latent Dirichlet Allocation (LDA)~\cite{Blei2003} topic modeling technique and benefit from the second-order semantic similarity metrics~\cite{Kolb2008,Kolb2009}.

In summary, our five key contributions are:
\begin{enumerate}
    \item the SARAv2 approach -- an extension of SARA~\cite{Palma2017SemanticAO} -- for the syntactic and semantic analysis of REST APIs for IoT applications;
    \item the definitions of three new linguistic patterns and antipatterns and their detection algorithms;
    \item an empirical assessment of the linguistic quality of a set of 19 IoT APIs from 18 different IoT providers;
    \item a web-based tool, \textsf{REST-Ling} available on \url{https://rest-ling.com} for the detection of linguistic and structural antipatterns and patterns;
    \item the empirical validation of the \textsf{REST-Ling} tool focusing on its accuracy and efficiency;
    \item a comparison with relevant studies on the detection of linguistic patterns and antipatterns from other domains (\textit{i.e.}, APIs for Cloud services and Web applications).
\end{enumerate}

We also perform a comparison with relevant studies on the detection of linguistic patterns and antipatterns from other domains (\textit{i.e.}, APIs for Cloud services and Web applications) from the perspectives of antipatterns and various accuracy measures. To assess the linguistic quality of IoT APIs and validate the SARAv2 approach, we define and answer the following four research questions:

\begin{itemize}
    \item RQ$_1$ Prevalence: \textit{To what extent IoT APIs suffer from poor linguistic design quality, i.e., linguistic antipatterns?}
    
    \item RQ$_2$ Comparison: \textit{To what extent APIs across domains suffer from poor linguistic design quality, i.e., linguistic antipatterns?}
    
    \item RQ$_3$ Accuracy: \textit{What is the accuracy of \textsf{REST-Ling} on the detection of linguistic antipatterns?}
    
    \item RQ$_4$ Efficiency: \textit{How does the \textsf{REST-Ling} perform in terms of average detection time for linguistic antipatterns?}
    
\end{itemize}

Our empirical results show that (1) out of the 19 analysed IoT APIs, only a few of them have syntactic design problems and most of the analysed URIs follow good linguistic practices, although there also exist certain poor practices in some specific APIs. Examples include: \textcolor{red}{\begin{scriptsize}\ding{54}\end{scriptsize}}\textit{Non-pertinent Documentation} was common in all IoT APIs and majority of the APIs had \textcolor{red}{\begin{scriptsize}\ding{54}\end{scriptsize}}\textit{Unversioned URI} antipattern. In contrast, almost all of the APIs followed \textcolor{teal}{\begin{scriptsize}\ding{52}\end{scriptsize}}\textit{Tidy URI} and \textcolor{teal}{\begin{scriptsize}\ding{52}\end{scriptsize}}\textit{Consistent Documentation} patterns; and (2) the \textsf{REST-Ling} tool has an average accuracy over 80\% when analysing IoT APIs.


The remaining article is structured as follows: Section~\ref{sec:LAPs} describes the linguistic patterns and antipatterns studied. Section~\ref{Approach} presents the SARAv2 approach we apply for the detection of linguistic patterns and antipatterns. Section~\ref{sec:experiments} shows experimental details and discusses the obtained detection results. Section~\ref{sec:related} discusses related works and makes a comparison with other state-of-the-art studies. Finally, in Section~\ref{Conclusion and Future Work} we conclude the research and present future work.

\section{Linguistic Patterns and Antipatterns}\label{sec:LAPs}

In total, we gathered nine linguistic patterns and antipatterns. The first six patterns and antipatterns are from the literature on REST APIs~\cite{Masse2012,Palma2015AreRA,RESTfulBestPractices,Palma2017SemanticAO,Berners-Lee2005,tilkovantipatterns2008} and the final three antipatterns are newly defined in this study.


To define new linguistic antipatterns, we studied similar linguistic antipatterns that exist in the object-oriented literature (e.g., related to class or method signature and source code comments), and performed a data analysis by also looking at the URIs and API documentation in our API dataset, to see the applicability of those linguistic antipatterns, and created "themes" of patterns and antipatterns that are applicable to REST URIs and documentation, by using thematic analysis \cite{Braun2006}. We adapted the detection heuristics from the object-oriented domain to the context of APIs that have resource identifiers (i.e., URIs) and their documentation. We defined \textcolor{red}{\begin{scriptsize}\ding{54}\end{scriptsize}}\textit{Inconsistent Documentation} linguistic antipattern being inspired from \cite{Arnaoudova2016}. We also studied the gray literature to discover concerns from the practitioners and formalise those observations in the form of linguistic antipatterns and their corresponding patterns. For example, the concept of \textcolor{red}{\begin{scriptsize}\ding{54}\end{scriptsize}}\textit{Unversioned URI} antipattern was discussed in~\cite{RESTAPIVersioning}. Another newly defined antipattern \textcolor{red}{\begin{scriptsize}\ding{54}\end{scriptsize}}\textit{Non-standard URI Design} is defined based on the notion similar to \textcolor{red}{\begin{scriptsize}\ding{54}\end{scriptsize}}\textit{Amorphous URI} antipattern (which affects the readability of the URIs) that non-standard characters should not be used in the URI design.

We formulated the detection heuristics of new linguistic antipatterns and patterns after a thorough discussion with the team consisting of two authors (who are not part of the manual validation). In the case of disagreement between the authors, a third opinion was sought from a researcher who also is not part of the experiment and validation. This enabled us to resolve the conflicts and avoid the bias by a specific author in defining new linguistic antipatterns and their detection heuristics. These new patterns and antipatterns are also applicable to APIs for Web applications or cloud services. The following subsections summarise the linguistic patterns and antipatterns SARAv2 can detect in REST APIs.

\subsection{Tidy \textit{vs.} Amorphous URIs}

The URIs in REST should be tidy and easy to read. A \textit{Tidy URI} has an appropriate lower-case resource naming, no extensions, underscores, or trailing slashes. \textit{Amorphous URI} occurs when URIs contain symbols or capital letters that make them difficult to read and use. A URI is amorphous if it contains: (1) upper-case letter (except for Camel Cases~\cite{MicrosoftMSDN}), (2) file extensions, (3) underscores, and, (4) a final trailing-slash~\cite{Masse2012, Palma2015AreRA}. The URI \url{www.exampleAlbum.com/NEW_Customer/ image01.tiff/} is a \textcolor{red}{\begin{scriptsize}\ding{54}\end{scriptsize}}\textit{Amorphous URI} since it includes a file extension, upper-case resource names, underscores, and a trailing slash. In contrast, the URI \url{www.example.com/customers/1234} is a \textcolor{teal}{\begin{scriptsize}\ding{52}\end{scriptsize}}\textit{Tidy URI} since it only contains lower-case resource naming, without extensions, underscores, or trailing slashes. The detection of this design practice requires syntactic analysis of the URIs.


\subsection{Contextualised \textit{vs.} Contextless Resource Names}

URIs should be \textit{contextual}, \textit{i.e.}, nodes in URIs should belong to semantically-related context. Thus, the \textit{Contextless Resource Names} appears when URIs are composed of nodes that do not belong to the same semantic context \cite{RESTfulBestPractices}. The URI \url{www.example.com/newspapers/planet/players?id=123} is a \textcolor{red}{\begin{scriptsize}\ding{54}\end{scriptsize}}\textit{Contextless Resource Names} because 'newspapers', 'planet', and 'players' do not belong to same semantic context. In contrast, the URI 
\url{www.example.com/soccer/team/players?id=123} is a \textcolor{teal}{\begin{scriptsize}\ding{52}\end{scriptsize}}\textit{Contextual Resource Names} because 'soccer', 'team', and 'players' belong to same semantic context. The detection of \textit{Contextualised vs. Contextless Resource Names} requires semantic analysis of the URIs.


\subsection{Verbless \textit{vs.} CRUDy URIs}

Appropriate HTTP methods, \textit{e.g.}, GET, POST, PUT, or DELETE, should be used in \textit{Verbless URIs} instead of using CRUDy terms (\textit{e.g.}, create, read, update, delete, or their synonyms) \cite{RESTfulBestPractices}. The use of such terms as resource names or requested actions is highly discouraged \cite{Masse2012, RESTfulBestPractices}. This URI with the HTTP POST \url{www.example.com/update/players/age?id=123} is a \textcolor{red}{\begin{scriptsize}\ding{54}\end{scriptsize}}\textit{CRUDy URIs} since it contains a CRUDy term 'update' while updating the user's profile color relying on an HTTP POST method. In contrast, this URI with the HTTP method POST \url{www.example.com/players/age?id=123} is a \textcolor{teal}{\begin{scriptsize}\ding{52}\end{scriptsize}}\textit{Verbless URIs} making an HTTP POST request without any verb. The detection of this design practice requires semantic analysis of the URIs.


\subsection{Hierarchical \textit{vs.} Non-hierarchical Nodes}

Nodes in a URI should be hierarchically related to its neighbor nodes. In contrast, \textit{Non-hierarchical Nodes} is an antipattern that appears when at least one node in a URI is not hierarchically related to its neighbor nodes \cite{RESTfulBestPractices}. The URI \url{www.examples1.com/professors/faculty/university} is a \textcolor{red}{\begin{scriptsize}\ding{54}\end{scriptsize}}\textit{Non-hierarchical Nodes}  since 'professors', 'faculty', and 'university' are not in a hierarchical relationship. In contrast, the URI \url{www.examples2.com/university/faculty/professors} is a \textcolor{teal}{\begin{scriptsize}\ding{52}\end{scriptsize}}\textit{Hierarchical Nodes} since 'university', 'faculty', and 'professors' are in a hierarchical relationship. The detection of \textit{Hierarchical vs. Non-hierarchical Nodes} requires semantic analysis of the URIs.


\subsection{Singularised \textit{vs.} Pluralised Nodes}

URIs should use singular/plural nouns consistently for resources naming across the API. When clients send PUT or DELETE requests, the last node of the request URI should be singular. In contrast, for POST requests, the last node should be plural. Therefore, the \textit{Pluralised Nodes} antipattern appears when plural names are used for PUT/DELETE requests or singular names are used for POST requests. However, GET requests are not affected by this antipattern~\cite{RESTfulBestPractices, Palma2015AreRA}. The first example URI is a POST method that does not use a pluralised resource, thus leading to \textcolor{red}{\begin{scriptsize}\ding{54}\end{scriptsize}}\textit{Pluralised Nodes}. In contrast, in the second example as shown below, for the \textcolor{teal}{\begin{scriptsize}\ding{52}\end{scriptsize}}\textit{Singularised Nodes}, the DELETE request acts on a single resource for deleting it. An example of \noindent\textcolor{red}{\begin{scriptsize}\ding{54}\end{scriptsize}}\textit{Pluralised Nodes} is DELETE \url{www.example.com/team/players} or POST \url{www.example.com/team/player}. The \noindent\textcolor{teal}{\begin{scriptsize}\ding{52}\end{scriptsize}}\textit{Singularised Nodes} can be exemplified as DELETE \url{www.example.com/team/player} or POST \url{www.example.com/team/players}. The detection of this design practice requires semantic analysis of the URIs.


\subsection{Pertinent \textit{vs.} Non-pertinent Documentation}

The \textcolor{red}{\begin{scriptsize}\ding{54}\end{scriptsize}}\textit{Non-pertinent Documentation} occurs when the documentation of a REST resource URI is in contradiction with its structure (\textit{e.g.}, nodes separated by slashes in URIs), inspired from a similar antipattern from the OO domain~\cite{Arnaoudova2016}. This antipattern applies to both a resource URI and its corresponding documentation. In contrast, a well-documented URI should properly and clearly describe its purpose using semantically related terms \cite{Arnaoudova2016,Petrillo10.1007978-3-319-94959-816}. The URI-documentation pair from Twitter:  \url{api.twitter.com/1.1/favorites/list} -- \textit{`Returns the 20 most recent Tweets liked by the authenticating or specified user'} shows no semantic similarity between them and, thus, considered as a \textcolor{red}{\begin{scriptsize}\ding{54}\end{scriptsize}}\textit{Non-pertinent Documentation}. In contrast, this URI-documentation pair from Instagram: \url{instagram.com/media/media-id/comments} -- `\textit{Gets a list of recent comments on a media object. The public content permission scope is required to get comments for a media that does not belong to the owner of the access token.}' shows a high relatedness and considered as a \textcolor{teal}{\begin{scriptsize}\ding{52}\end{scriptsize}}\textit{Pertinent Documentation}. The detection of this design practice requires semantic analysis of the URIs and their documentations.



\subsection{Consistent \textit{vs.} Inconsistent Documentation}

The \textcolor{red}{\begin{scriptsize}\ding{54}\end{scriptsize}}\textit{Inconsistent Documentation} found in REST API documentation is defined based on another antipattern \textit{Method Signature and Comment are Opposite}~\cite{Arnaoudova2016} common in object-oriented systems. It occurs if the documentation of a method is in contradiction with its declaration. REST API documentations may also manifest similar practice where a resource URI (together with the HTTP method) is in contradiction with its documentation. For example, in the IBM Watson IoT, the POST method with \url{/bulk/devices/remove} URI is in contradiction with its documentation '\textit{Delete multiple devices. Delete multiple devices, each request can contain a maximum of 512kB}', thus, is an \textcolor{red}{\begin{scriptsize}\ding{54}\end{scriptsize}}\textit{Inconsistent Documentation}. By REST design principles, the POST method should be used to create something. When a resource URI (together with the HTTP method) is in contradiction with its documentation. For the same example URI, \url{/bulk/devices/remove}, if the  documentation were stated as '\textit{Remove multiple devices. Remove multiple devices, each request can contain a maximum of 512kB}', this could be identified as \textcolor{teal}{\begin{scriptsize}\ding{52}\end{scriptsize}}\textit{Consistent Documentation}. The detection of \textit{Consistent vs. Inconsistent Documentation} requires semantic analysis of the URIs and their documentations.


\subsection{Versioned \textit{vs.} Unversioned URIs}

APIs evolve continuously and if not properly versioned, might cause clients to break. Versioned APIs facilitates easy maintenance both for API providers and client developers. Changes in the APIs may include a change in the response data format or type, removing a resource, adding a new end-point, response parameters, which require to track major or minor versions for APIs. If an API is not versioned at all, the \textit{Unversioned URI} linguistic antipattern occurs~\cite{RESTAPIVersioning}. For example, Losant API does not have any version information in its URI, whereas IBM Watson IoT always use versioned URI, thus follow the \textit{Versioned URI} pattern \cite{RESTAPIVersioning}. The URI \url{api.example.com/1.1/resourceid/view} is an example of \textcolor{teal}{\begin{scriptsize}\ding{52}\end{scriptsize}}\textit{Versioned URI} with its API version embedded in the URI. In contrast, the URI \url{api.example.com/resourceid/view} is an example of \textcolor{red}{\begin{scriptsize}\ding{54}\end{scriptsize}}\textit{Unversioned URI}. The detection of this design practice requires syntactic analysis of the URIs.

\subsection{Standard vs. Non-standard URI}\label{nonstandardURI}

The URI design should not include nodes or resources with non-standard identification, which hinders the reusability and understandability of the APIs. The \textcolor{red}{\begin{scriptsize}\ding{54}\end{scriptsize}}\textit{Non-standard URI Design} occurs when (1) characters like é, å, ö, etc. are present in URIs, (2) blank spaces are found in URIs, (3) double hyphens are used in URIs, and (4) unknown characters (\textit{e.g.}, !, @, \#, \$, \%, \^{}, \&, *, etc.) are present in URIs. Instead, a URI following \textcolor{teal}{\begin{scriptsize}\ding{52}\end{scriptsize}}\textit{Standard URI Design} (1) does not include non-standard characters like é, å, ö, etc. and (2) replaces blank spaces, unknown characters, and double hyphens with a single hyphen. The URI \texttt{api.example.com/museum/louvre/réception/} is an example of \textcolor{red}{\begin{scriptsize}\ding{54}\end{scriptsize}}\textit{Non-standard URI Design}. While, the URI \texttt{api.example.com/museum/louvre/reception/} represents \textcolor{teal}{\begin{scriptsize}\ding{52}\end{scriptsize}}\textit{Standard URI Design}. The first example format hinders the usability and understandability as compared to the latter URI. The detection of \textit{Standard vs. Non-standard URI} requires syntactic analysis of the URIs.

\section{The SARAv2 Approach}\label{Approach}

\begin{figure}[t!]
    \centering
		\includegraphics[width=.95\textwidth]{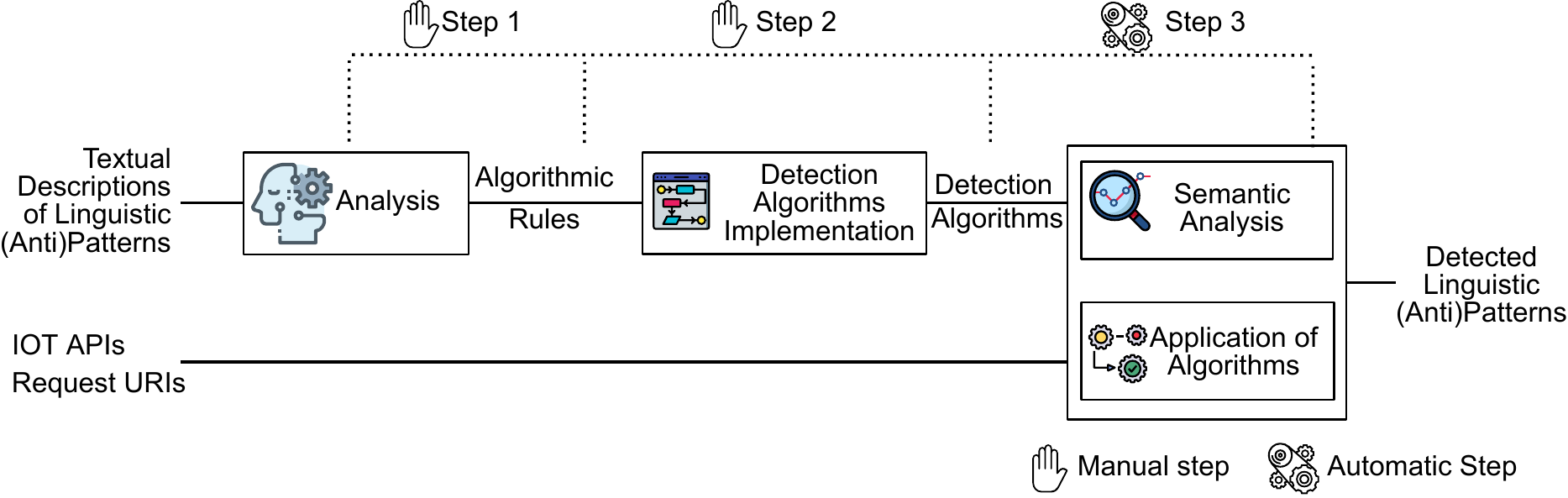}
    \caption{Overview of the SARAv2 approach.} \label{Implementation details}
\end{figure}

SARAv2 (Semantic Analysis of REST APIs version two) enables the automatic detection of nine linguistic antipatterns and their corresponding patterns in REST APIs and, therefore, in REST APIs for IoT. To analyse the REST APIs and their documentation, we manually collect a subset of URIs and their documentation provided by each REST API provider (\textit{i.e.}, in this paper IoT providers). These collected URIs and their documentation are used later in the detection phase. As shown in Figure \ref{Implementation details}, the SARAv2 approach consists of three steps:

\vspace{+1mm}
\noindent\textit{Step 1. Analysis of Linguistic Patterns and Antipatterns:} A manual step that consists of analysing the description of linguistic patterns and antipatterns from the literature to identify the properties relevant to their detection. We use these relevant properties to define \textit{detection heuristics} for patterns and antipatterns.

\vspace{+1mm}
\noindent\textit{Step 2. Implementation of Detection Algorithms:} A second manual step that involves the implementation of concrete detection algorithms for patterns and antipatterns based on the detection heuristics defined in Step 1.

\vspace{+1mm}
\noindent\textit{Step 3. Detection of Linguistic Patterns and Antipatterns:} An automatic step that executes the semantic analysis of resource URIs and API documentation by automatically applying the detection algorithms (implemented in Step 2) on URIs and APIs documentation for the detection of linguistic patterns and antipatterns.

In the following sections, we discuss each step in SARAv2 in detail.

\subsection{Analysis of Linguistic Patterns and Antipatterns}\label{sec:Analysis of Linguistic Patterns and Antipatterns}

We analyse the definitions of antipatterns and patterns defined in Section \ref{sec:LAPs} to identify their various linguistic aspects. For example, a linguistic aspect for the detection of the \textcolor{red}{\begin{scriptsize}\ding{54}\end{scriptsize}}\textit{Contextless Resource Names} is to assess whether a pair of URI nodes are semantically related, \textit{i.e.}, belong to the same semantic context. Figure \ref{Algorithmic rule of Contextless Resource Names.} shows the detection heuristic for the \textcolor{red}{\begin{scriptsize}\ding{54}\end{scriptsize}}\textit{Contextless Resource Names}. We extract the domain knowledge from the URI documentation and the request URI (lines 2-3), to build a topic model. We then calculate and check the similarity among the nodes (line 4). We check this similarity using the topic model we generate by applying natural language processing techniques. We calculate the average similarity value for all the nodes in a URI against each topic from our topic model. And, we report a URI has the \textcolor{red}{\begin{scriptsize}\ding{54}\end{scriptsize}}\textit{Contextless Resource Names} if the average similarity value is less than the threshold (line 5). On the contrary, an occurrence of \textcolor{teal}{\begin{scriptsize}\ding{52}\end{scriptsize}}\textit{Contextual Resource Names} will be reported if the similarity value is equal to or higher than the threshold. Based on our previous studies~\cite{Palma2017SemanticAO} and findings by Kolb~\cite{Kolb2009}, we used 0.3 as the threshold to determine semantic relatedness between words. We empirically determined the threshold value, \textit{i.e.}, we started from 0.1 and increased 0.05 each time the semantic relatedness for a set of pair of nodes is not reasonable. Moreover, on using DISCO, Kolb~\cite{Kolb2008, Kolb2009} determined the threshold value of 0.3 can be utilised as the \textit{gold standard} with good accuracy with regard to semantic relatedness.

\begin{figure}[ht!]
\begin{flushleft}
\ttfamily
\scriptsize
\hbox{\raisebox{0.05em}{\vrule depth 0pt height 0.25pt width \columnwidth} }
1: \textsc{Contextless-Resource}(\textit{Request-URI, API-Documentation})\\
2: \hspace{1.25em}\textit{TopicsModel} $\gets$ \textsc{Extract-Topics}(\textit{API-Documentation})\\
            3: \hspace{1.25em}\textit{URINodes} $\gets$ \textsc{Extract-URI-Nodes}(\textit{Request-URI})\\
4: \hspace{1.25em}\textit{Similarity-Value} $\gets$\textsc{Calculate-Second-Order-Similarity}(\textit{URINodes}, \textit{TopicsModel})\\
5: \hspace{1.25em}\textbf{if} \textit{Similarity-Value} $<$ \emph{threshold}\\
6: \hspace{1.25em}\hspace{1.25em}\textbf{return} \textit{true ``Contextless Resource Names" antipattern}\\
7: \hspace{1.25em}\textbf{end if}\\
8: \hspace{1.25em}\textbf{return} \textit{false ``Contextual Resource Names" pattern}\\
\hbox{\raisebox{0.05em}{\vrule depth 0pt height 0.25pt width \columnwidth} }
\caption{Detection heuristic for \textit{Contextless Resource Names} antipattern.}
\label{Algorithmic rule of Contextless Resource Names.}
\end{flushleft}
\end{figure}

Similarly, a linguistic aspect for the detection of the \textcolor{red}{\begin{scriptsize}\ding{54}\end{scriptsize}}\textit{Inconsistent Documentation} is to assess whether the HTTP method used with a resource URI is described with a conflicting documentation. Figure \ref{Algorithmic rule of REST antipattern LDA.} presents the detection heuristic for the \textcolor{red}{\begin{scriptsize}\ding{54}\end{scriptsize}}\textit{Inconsistent Documentation}. We begin with preprocessing the documentation by removing the stop words (line 2) and tokenise the documentation, i.e., obtain the set of words in the documentation, and lemmatise them, \textit{i.e.}, we extract the base form of each word in the documentation  (line 3). Then, we match with the HTTP method to check if the URI and its documentation are related and consistent (lines 4 to 13). To measure relatedness between the HTTP method and the documentation, we check whether the synonyms of various actions or verbs are misplaced within the documentation. For example, the HTTP POST method is often used to create a new resource if the resource does not exist already. Thus, the documentation related to this action or the resource on which the POST action is taken, must not have any indication of resource deletion, retrieval, or update. If such contradiction is found in the documentation then SARAv2 will report a \textcolor{red}{\begin{scriptsize}\ding{54}\end{scriptsize}}\textit{Inconsistent Documentation} (lines 4-5). In contrast, an occurrence of \textcolor{teal}{\begin{scriptsize}\ding{52}\end{scriptsize}}\textit{Consistent Documentation} will be reported if no contradiction is discovered between the HTTP method/action and the documentation. The detection heuristics of other linguistic patterns and antipatterns are available online\footnote{\url{http://sofa.uqam.ca/santa/}}.

\begin{figure}[ht!]
\begin{flushleft}
\ttfamily
\scriptsize
\hbox{\raisebox{0.05em}{\vrule depth 0pt height 0.25pt width \columnwidth} }
1: \textsc{Inconsistent-Documentation}(\textit{HTTP-Method, Request-URI}, \textit{Documentation})\\
2: \hspace{1.35em}\textit{Documentation} $\gets$ \textsc{Remove-Stop-Words}(\textit{Documentation})\\
3: \hspace{1.35em}\textit{Tokens} $\gets$ \textsc{Lemmatise-Tokenise}(\textit{Documentation})\\
4: \hspace{1.35em}\textbf{if} \textit{HTTP-Method} = `POST' \textbf{\&\&} \textsc{Synonyms}(Delete or Update or Get) $\in$ Tokens\\
5: \hspace{1.35em}\hspace{1.25em}\textbf{return} \textit{true ``Inconsistent Documentation" antipattern}\\
6: \hspace{1.35em}\textbf{else if} \textit{HTTP-Method} = `DELETE' \textbf{\&\&} \textsc{Synonyms}(Create or Update or Get) $\in$ Tokens\\
7: \hspace{1.35em}\hspace{1.25em}\textbf{return} \textit{true ``Inconsistent Documentation" antipattern}\\
8: \hspace{1.35em}\textbf{else if} \textit{HTTP-Method} = `PUT' \textbf{\&\&} \textsc{Synonyms}(Create or Delete or Get) $\in$ Tokens\\
9: \hspace{1.35em}\hspace{1.25em}\textbf{return} \textit{true ``Inconsistent Documentation" antipattern}\\
10: \hspace{1em}\textbf{else if} \textit{HTTP-Method} = `GET' \textbf{\&\&} \textsc{Synonyms}(Delete or Update or Create) $\in$ Tokens\\
11: \hspace{1em}\hspace{1.25em}\textbf{return} \textit{true ``Inconsistent Documentation" antipattern}\\
12: \hspace{1em}\textbf{end if}\\
13: \hspace{1em}\textbf{return} \textit{false ``Consistent Documentation" pattern}\\
\hbox{\raisebox{0.05em}{\vrule depth 0pt height 0.25pt width \columnwidth} }
\caption{Detection heuristic for \textit{Inconsistent Documentation} antipattern.}
\label{Algorithmic rule of REST antipattern LDA.}
\end{flushleft}
\end{figure}

\subsection{Implementation of Detection Algorithms}

To detect the linguistic pattern and antipattern, we implemented the detection algorithms using Java since our detection framework, SOFA (Service Oriented Framework for Antipatterns)~\cite{Palma2017SemanticAO}, is Java-based. The SARAv2 approach does not require the parameterised URIs to perform the analysis, \textit{i.e.}, performs the analysis on the URIs from the IoT APIs documentation. We manually convert (write the Java code) the detection heuristics defined in Section~\ref{sec:Analysis of Linguistic Patterns and Antipatterns} into executable Java programs. Listing 1 shows an example of a code snippet in the form of pseudocode that we apply for the detection of \textcolor{red}{\begin{scriptsize}\ding{54}\end{scriptsize}}\textit{Contextless Resource Names}.

As shown in Listing 1, once the \texttt{detectContextlessResource()} method is invoked (line 4) the \texttt{URIContextualAnalysis()} procedure is initiated (line 9), which is the implementation of the heuristics for \textit{Contextless Resource Names} antipattern in Figure \ref{Algorithmic rule of Contextless Resource Names.}. Inside the \texttt{URIContextualAnalysis()} procedure, first, the topic model is built (line 18), followed by the extraction of the nodes in the URI (line 20). We use a matrix for storing the similarity values between each node and the members in the topic model (line 22). Finally, the calculation of second-order similarity values takes place (lines 24 and 25), and the detection of either an antipattern or a pattern is decided based on the average similarity value for each node. If the average similarity value for all the nodes in a URI is below a predefined threshold (lines 27 to 30), we consider it as a contextless URI design, thus, an antipattern, and if above, it is considered as a contextual URI design, \textit{i.e.}, a pattern.

\lstdefinestyle{interfaces}{
  float,
  floatplacement=tbp
}
\lstset{showstringspaces=false, numbersep=1.5pt, breaklines=true, xleftmargin=1.5pt, xrightmargin=1pt, keywordstyle=\color{blue}, numbers=left, stepnumber=1, language=Java, caption=Code snippet for the detection \textit{Contextless Resource Names} linguistic antipattern.}
\begin{lstlisting}[style=interfaces, basicstyle=\scriptsize]
public static void main(String[] args) {
    ...
    /* Detection of Contextless Resource Names Antipattern */
    detectContextlessResourceNames();
    writeOutput();
}
private static void detectContextlessResource() {	
    result = restAnalyser.URIContextualAnalysis(URI);
    if(!result)
        contextlessResultAP.add(URI);
    else
        contextlessResultP.add(URI);
}
public boolean URIContextualAnalysis(String Uri) {
    ...
    Topics = this.ldaProcessor.getTopicList();
    ...
    ArrayList<String> UriNodes = getUriNodes(Uri);
    ...
    for each Node in UriNodes and for each topic in Topics;
        Calc_Second_Order_Sim_Values();
    Calc_Avg_Sim_of_Nodes_in_URI();
    if average_similarity < threshold
        return true;
    else
        return false;
}	
\end{lstlisting}

\subsection{Detection of Linguistic Patterns and Antipatterns}

In SARAv2, the detection of linguistic patterns and antipatterns utilises two essential elements: the Second Order Semantic Similarity metric and Latent Dirichlet Allocation (LDA).

The Second Order Semantic Similarity metric~\cite{Kolb2008, Kolb2009} allows obtaining the distributionally most similar words for a given word, and computes similarity scores among them based on second-order word vectors. Two words are considered distributionally similar if they have multiple co-occurring words in the same syntactic relations~\cite{Budanitsky2006}. The distributional semantic similarity goes beyond \textit{is-a} relationships between nouns and verbs as allowed by approaches~\cite{Budanitsky2006,Kolb2009} based on WordNet~\cite{wordnet} that only benefits from the synonym (warm-hot), meronym (car-wheel), and antonym (hot-cold) relations. Distributional semantic similarity captures the multiple senses of a given word and allows mixing all the distributionally similar semantic words for all these senses.

LDA is a generative probabilistic model of a corpus based on topic models. It relies on the idea that a document is a mixture of latent topics, and each topic is a probabilistic distribution over words~\cite{Blei2003} and supports the extraction of topic models from a corpus. The topic model is a low-dimensional representation of the content of the documents in the corpus. LDA allows a document to pertain to many different topics by associating the probability of the document belonging to each topic, overcoming one of the main problems of many other clustering models that restrict documents to be associated with just one topic. LDA is also affected by the \emph{bag-of-words} assumptions, meaning that words that appear or should be generated by a topic might also be allocated in other topics \cite{Blei2003}. To tackle these problems, we defined a hybrid approach for SARA~\cite{Palma2017SemanticAO}, combining LDA topic modeling to obtain the low-dimensional representation of the corpus and the distributional semantic similarity to measure the semantic similarity between the words.

\begin{figure}[t!]
    \centering
        \includegraphics[width=.9\textwidth]{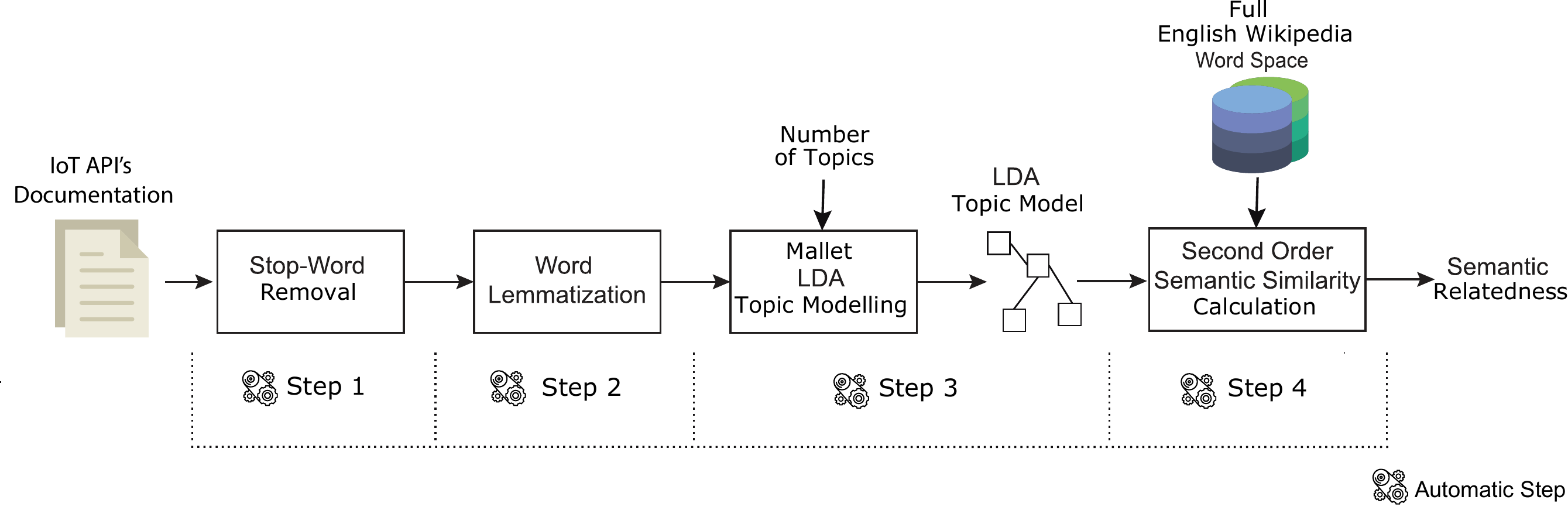}
    \caption{The semantic analysis strategy of the SARAv2 approach.} \label{Semantic Analysis}
\end{figure}

\subsubsection{Semantic Analysis of IoT APIs}

SARAv2 strategy for performing the semantic analysis of IoT APIs involves four automatic steps, as illustrated in Figure \ref{Semantic Analysis}: (1) collecting IoT APIs documentation and their preprocessing (\textit{i.e.}, exclusion of stop words); (2) truncating example URI nodes to their base form (\textit{i.e.}, lemmatisation) using Stanford's CoreNLP \cite{manning-EtAl:2014:P14-5}; (3) extracting the LDA topic model using the collected corpora; and (4) measuring the second-order similarity between the extracted LDA topic model and the URI nodes. The LDA topic model is created by using the Mallet LDA topic modeling tool-set\footnote{http://mallet.cs.umass.edu} using the IoT APIs documentation, excluding lists of parameters, and response formats. This LDA model represents a minimal representation of the members of the corpus, preserving the essential semantic relationships needed for classification~\cite{Blei2003}. 

In the following, we briefly describe how we determine the detection of a linguistic antipattern \textit{Contextless Resource Names} (and the corresponding \textit{Contextual Resource Names} pattern) using the LDA topic modeling~\cite{Blei2003} and second-order semantic similarity~\cite{Kolb2008, Kolb2009}.

\begin{table}[htb!]
\footnotesize
\centering
\def\arraystretch{.75}
\setlength\tabcolsep{5pt}
\begin{tabular}{ccc}
\toprule
\textbf{Topic 1} & \textbf{Topic 2} & \textbf{Topic 3}\\
\midrule
eco	& smoke & device \\
record & sound & structure \\
estimate & snapshot & thermostat \\
lock & status & event \\
adjust & change & nest \\
format & list & camera \\
related & display & url \\
json & nest & display \\
call & expire & temperature \\
sign & home & alarm \\
home & detect & require \\
sound & subscription & hvac \\
bandwidth & field & aware \\
low & live & zone \\
image & motion & activity \\
\bottomrule
\end{tabular}
\caption{Top 15 words for Google Nest topic model with k=3.}
\label{tab:topics}
\end{table}

\begin{table}[t!]
\centering
\caption{An example analysis of two Google Nest URIs.}\label{tab:topic model}
\def\arraystretch{0.9}
\setlength\tabcolsep{1.5pt}
\scriptsize
\begin{tabular}{|l|l|ccc|ccc|}
\hline
   &    \textbf{Topic}    & \multicolumn{3}{c|}{\begin{tiny}\textbf{/devices/thermostats/device\_id/locale}\end{tiny}} & \multicolumn{3}{c|}{\begin{tiny}\textbf{/structures/structure\_id/co\_alarm\_state}\end{tiny}} \\
\cline{3-8}
& \textbf{words} & \textit{device} & \textit{thermostat}  & \textit{locale} & \textit{structure} & \textit{alarm} & \textit{state} \\ 
\cline{1-8}

\multirow{15}{*}{\textbf{Topic 1}} & eco & 0.0528  & 0.0130  & 0.0275  & 0.0207  & 0.0146  & \textbf{0.0064}   \\
 & record & 0.0068  & 0.0000  & 0.0000  & 0.0000  & 0.0129  & 0.0000 		\\
 & estimate & 0.0611  & 0.0192  & 0.0241  & 0.0928  & 0.0239  & 0.0000 		\\
 & lock & 0.2254  & \textbf{0.3378}  & 0.0055  & 0.1275  & 0.1955  & 0.0000 		\\
 & adjust & 0.2026  & 0.2575  & 0.0170  & 0.0721  & 0.1936  & 0.0000 		\\
 & format & 0.5311  & 0.0922  & 0.0451  & 0.1478  & 0.0990  & 0.0000 		\\
 & related & 0.0349  & 0.0129  & 0.0000  & 0.0327  & 0.0232  & 0.0000 		\\
 & json & 0.1962  & 0.0182  & 0.0253  & 0.0702  & 0.0313  & 0.0000 		\\
 & call & 0.0656  & 0.0081  & 0.0058  & 0.0648  & 0.0915  & 0.0000 		\\
 & sign & 0.0000  & 0.0000  & 0.0057  & 0.0000  & 0.0060  & 0.0058 		\\
 & home & 0.0000  & 0.0000  & 0.0703  & 0.0817  & 0.0000  & 0.0062 		\\
 & sound & 0.1645  & 0.0919  & 0.1588  & 0.1853  & 0.2277  & 0.0000 		\\
 & bandwidth & \textbf{0.7259}  & 0.2130  & 0.0317  & 0.0944  & \textbf{0.2545}  & 0.0000 		\\
 & low & 0.0517  & 0.0958  & 0.0113  & 0.0521  & 0.0636  & 0.0000 		\\
 & image & 0.3504  & 0.0788  & \textbf{0.1595}  & \textbf{0.5273}  & 0.1274  & 0.0000 		\\
 \cdashline{2-8}[.75pt/1pt]
		& \textbf{Average} & 	\multicolumn{3}{c|}{\textbf{0.4077}} & 	\multicolumn{3}{c|}{\textbf{0.2627}} 	        \\ \hline
\multirow{15}{*}{\textbf{Topic 2}} & smoke & 0.0054  & 0.1065  & 0.0058  & 0.0059  & 0.0609  & 0.0000 	\\	
 & sound & 0.1645  & 0.0919  & \textbf{0.1588}  & 0.1853  & 0.2277  & 0.0000 		\\
 & snapshot & 0.3066  & 0.0178  & 0.0395  & 0.0732  & 0.0863  & 0.0000 		\\
 & status & 0.0588  & 0.0135  & 0.1419  & 0.2592  & 0.0188  & 0.0291 		\\
 & change & 0.0976  & 0.0603  & 0.0805  & 0.3649  & 0.0852  & 0.0000 		\\
 & list & 0.0348  & 0.0069  & 0.0741  & 0.1001  & 0.0054  & \textbf{0.1672} 		\\
 & display & \textbf{0.7431}  & 0.1763  & 0.0842  & 0.2419  & 0.1894  & 0.0000 		\\
 & nest & 0.0054  & 0.0000  & 0.0495  & 0.0118  & 0.0000  & 0.0061 		\\
 & expire & 0.0442  & 0.0000  & 0.0060  & 0.0055  & 0.0072  & 0.0364 		\\
 & home & 0.0000  & 0.0000  & 0.0703  & 0.0817  & 0.0000  & 0.0062 		\\
 & detect & 0.3562  & 0.1291  & 0.0000  & 0.1933  & \textbf{0.3026}  & 0.0000 		\\
 & subscription & 0.2463  & 0.0113  & 0.0059  & 0.0180  & 0.0687  & 0.0000	\\ 	
 & field & 0.1114  & 0.0193  & 0.1001  & 0.1900  & 0.0117  & 0.0136 		\\
 & live & 0.0000  & 0.0000  & 0.0836  & 0.0260  & 0.0062  & 0.0069 		\\
 & motion & 0.3378  & \textbf{0.1946}  & 0.0657  & \textbf{0.4712}  & 0.2228  & 0.0000 		\\
 \cdashline{2-8}[.75pt/1pt]
		& \textbf{Average} & 	\multicolumn{3}{c|}{\textbf{0.3655}} & 	\multicolumn{3}{c|}{\textbf{0.3137}} 	        \\ \hline
\multirow{15}{*}{\textbf{Topic 3}} & device & \textbf{2.0000}  & 0.4916  & 0.0242  & 0.2111  & 0.4377  & 0.0000 \\
 & structure & 0.2111  & 0.0569  & 0.1645  & \textbf{2.0000}  & 0.0210  & 0.0000 \\
 & thermostat & 0.4916  & \textbf{2.0000}  & 0.0057  & 0.0569  & 0.3944  & 0.0000 \\
 & event & 0.0175  & 0.0118  & 0.1557  & 0.0507  & 0.0152  & 0.0000 \\
 & nest & 0.0054  & 0.0000  & 0.0495  & 0.0118  & 0.0000  & 0.0061 \\
 & camera & 1.0680  & 0.3072  & 0.0177  & 0.1089  & 0.4096  & 0.0000 \\
 & url & 0.2685  & 0.0117  & 0.0177  & 0.0577  & 0.0766  & 0.0000 \\
 & display & 0.7431  & 0.1763  & 0.0842  & 0.2419  & 0.1894  & 0.0000 \\
 & temperature & 0.0987  & 0.2084  & 0.0645  & 0.1731  & 0.0681  & 0.0000  \\  
 & require & 0.4377  & 0.3944  & 0.0000  & 0.0210  & 0.1611  & 0.0000 \\
 & alarm & 0.3914  & 0.1334  & 0.0167  & 0.1543  & \textbf{2.0000}  & 0.0140 \\
 & hvac & 0.3133  & 0.8992  & 0.0000  & 0.0351  & 0.2735  & 0.0000 \\
 & aware & 0.0060  & 0.0000  & 0.0000  & 0.0116  & 0.1789  & 0.0000 \\
 & zone & 0.0729  & 0.0402  & \textbf{0.4626}  & 0.2317  & 0.0195  & \textbf{0.0727} \\
 & activity & 0.1565  & 0.0371  & 0.2037  & 0.3848  & 0.0912  & 0.0053 \\
 \cdashline{2-8}[.75pt/1pt]
		& \textbf{Average} & 	\multicolumn{3}{c|}{\textbf{1.4875}} & 	\multicolumn{3}{c|}{\textbf{1.3576}}  \\
\hline
\end{tabular}
\end{table}

\subsubsection{Determining Patterns and Antipatterns}

To discover the relationships (\textit{e.g.}, contextual) between the pair of nodes (\textit{i.e.}, resource identifiers) in the URIs, as mentioned above, we use the tool-set based on Mallet LDA topic modeling. Given a collection of text (or documents), LDA generates a topic model that specifies the important relationships crucial for classification or summarisation tasks~\cite{Blei2003}. In other words, the generated topic model represents the collection of documents in some low-dimensional word vectors. The topic model for each IoT API was built after gathering the descriptions of the resource URIs as input by excluding the list of parameters, request or response formats, and example code.

We start processing the collection of text by removing the stop words and expanding the acronyms to get the full form. For this, we collect a list of API-specific acronyms. The collection of acronyms is performed after we gather 1,102 URIs and their documentation for 19 IoT APIs. We go through each URI and its documentation, look for acronyms and create an API-specific dictionary that includes the acronyms and their full forms. Later, during the processing of the URIs and their documentation, we replace the acronyms with their full forms to build a more accurate topic model. The lemmatisation process is also applied to set the words to their base form, for which we rely on Stanford CoreNLP~\cite{manning-EtAl:2014:P14-5}. Then, we obtain the topic model with \textit{k} topics using the Mallet tool-set for which the set of unique end-points for an IoT API is considered as topics. This is done because end-points are the key concepts for an API as they appear first in the URI design hierarchy~\cite{RESTfulBestPractices}.

Table \ref{tab:topics} shows the topic model obtained using Mallet from the documentation corpus of Google Nest IoT API. As shown, the topic model has three topics, and for each topic, the 15 most relevant words are listed. Later, we use this topic model to quantify the similarity between a pair of nodes in a URI. For example, two nodes (or resource identifiers) are related or similar if they belong to the same topic following the method proposed by Griffith and Steyvers~\cite{griffith2006probabilistic}.

After building the LDA topic model, we rely on the second-order semantic similarity metric to compute the (semantic or contextual) similarity between identifiers. The distributional second-order similarity metric is useful for us because the nodes (\textit{i.e.}, resource identifiers) might slightly differ from their actual API documentation syntactically and semantically. Two nodes (or words) can be seen as distributionally similar when they have co-occurring words in common, \textit{i.e.}, common words as neighbors. We rely on DISCO~\cite{Kolb2008} library to compute the distributional similarity between nodes in a URI.

Table \ref{tab:topic model} shows similarity values for two URIs from Google Nest API: (1) \url{developer-api.nest.com/devices/thermostats/device_id/locale} and (2) \url{developer-api.nest.com/structures/structure_id/co_alarm_state}. 

These values are computed based on the topic models and the distributional second-order similarity metric. In other words, if we want to compare the context of a pair of nodes in a URI, we compute the second-order semantic similarity between them with the top 15 words in each topic from the obtained topic model. Then, we decide the topic to which a node belongs to based on the similarity value, \textit{i.e.}, a node fits a topic if the average second-order semantic similarity value is greater than the threshold 0.3. Also, for a pair of nodes, if the intersection of topics to which the nodes belong is \textit{null} (\textit{i.e.}, no common topic), then, the URI is regarded as an instance of \textit{Contextless Resource Names} linguistic antipattern. In contrast, if each pair of nodes in a URI belongs to one or more common topic(s), we report the URI as an instance of \textit{Contextual Resource Names} linguistic pattern.

For the first URI, the base form of each node (\textit{i.e.}, \textit{device}, \textit{thermostat}, and \textit{locale}) appears in Topic 3, except the node \textit{locale}. Moreover, the average similarity value for all the nodes against Topic 1 is 0.4077, against Topic 2 is 0.3655, and against Topic 3 is 1.4875, which means the first URI is more similar to Topic 3 with a higher similarity value of 1.4875. The average similarity in Table \ref{tab:topic model} was computed by taking the maximum similarity value for each node in the URI against all the words in a topic, and then average them. For example, the maximum similarity values for the nodes in the first URI are 0.7259, 0.3378, and 0.1595 for Topic 1, averaging 0.4077, which is greater than the threshold of 0.3. The average similarity values for Topic 2 and Topic 3 are 0.3655 and 1.4875, respectively. Ergo, we identify the first URI as \textit{Contextual Resource Names} linguistic pattern. For the second Google Nest URI, all the nodes \textit{structure}, \textit{alarm}, and \textit{state} appear in Topic 3 in their base form except the node \textit{state}. Similar to the first URI, the second URI also more fit Topic 3 with an average similarity value of 1.3576 (see Table \ref{tab:topic model}). We identify the second URI as \textit{Contextual Resource Names} linguistic pattern because all the nodes are semantically related.

\subsubsection{Applying Detection Algorithms}

In this work, we use SARAv2, which extends the SARA approach for the semantic analyses of REST URIs and APIs documentations used in ~\cite{Palma2015AreRA,Palma2017SemanticAO}
by adding three new patterns and antipatterns. The  extension includes the implementation of the new detection algorithms for the three newly defined patterns and antipatterns.

Both SARA and SARAv2 use and extend the SOFA framework proposed and developed by Moha \textit{et al.}~\cite{10.1007/978-3-642-34321-61} to automatically execute the detection heuristics in the form of detection algorithms on the URIs. SARA and SARAv2 extend SOFA by enabling the use of LDA models and Second Order Semantic Similarity as heuristics in the detection algorithms that analyze the URIs and their documentation. For example, the detection code, as shown in Listing 1, is implemented and executed inside the SOFA framework. The detection results are then exported to a text file.




\subsection{Operationalising \textsf{REST-Ling} Tool}
\begin{figure}[t!]
\includegraphics[width=1\textwidth]{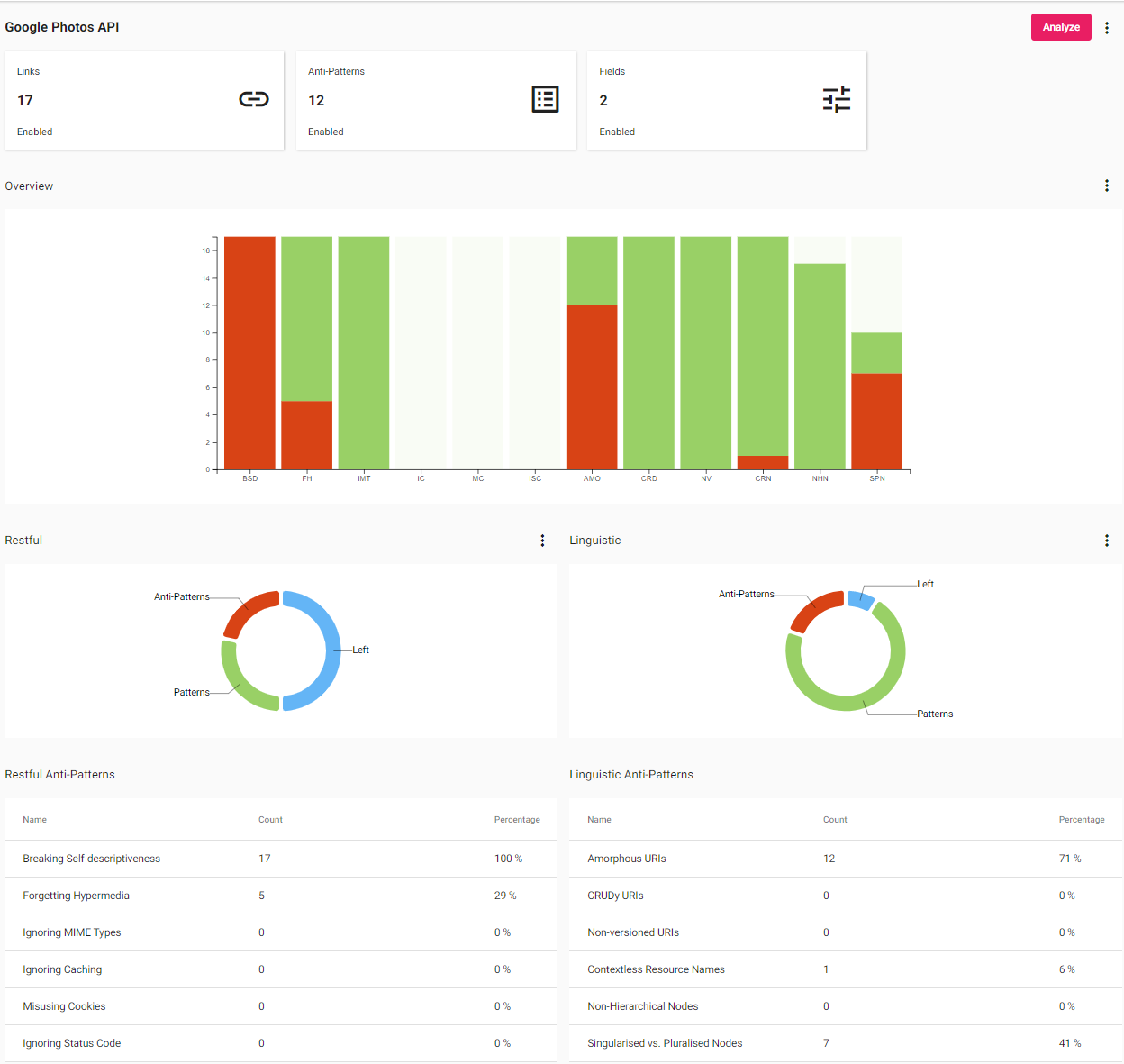}
\caption{Detection of patterns and antipatterns in an API.} \label{rest_ling_results}
\end{figure}

\textsf{REST-Ling} is a web application that automates the detection of linguistic patterns and antipatterns. \textsf{REST-Ling} aims to help software engineers analyse their APIs and detect linguistic patterns and antipatterns. The tool can present various visual representations for patterns and antipatterns detected in a particular API. Moreover, it shows generic information on the types of linguistic patterns and antipatterns detected, and pinpoints the rationale behind their detection. Our \textsf{REST-Ling} tool supports:

\begin{itemize}
\item \textit{The addition of APIs and URIs}: \textsf{REST-Ling} allows the user to add one or more URIs manually or by uploading a JSON file. The JSON file can contain multiple APIs with multiple URIs each; 
\item\textit{The selection of patterns and antipatterns}: \textsf{REST-Ling} allows the user to select the patterns and antipatterns (both design and linguistic) to be detected. The analysis process is done asynchronously for all the patterns and antipatterns; 
\item \textit{A detailed view of the detection results}: The tool provides answers to what, why, and where the design and linguistic antipatterns occur. This allows the user to have a better insight into the API quality by checking what type of antipatterns an API has; 
\item\textit{A graphical representation of the detection results}: The tool provides a graphical representation of the detection results of the patterns and antipatterns using pie and bar charts; 
\item\textit{A topic model creation feature for the linguistic analysis}: When it comes to detecting, for example, \textit{Contextless Resource Names} linguistic antipattern, the tool provides functionalities to import and add acronyms and stop words to create the topic model required for the detection.

\end{itemize}

\vspace{3mm}
\noindent\textbf{Example Use:} To use the \textsf{REST-Ling}, engineers require a JSON file that contains a list of URIs from an API. Each URI should have a name, method, and description. Users can upload the JSON file on the \texttt{Collections} page to add the APIs to be analysed. Once the file is uploaded, the user can go into each \texttt{Collection} and start the analysis by clicking on the \texttt{Analyse} button and checking the detected patterns and antipatterns in the collection view, as shown in Figure \ref{rest_ling_results}. The \textsf{REST-Ling} tool can be accessed on \url{https://rest-ling.com}. To use the tool, provide \texttt{admin} both as the username and password. A demo of the tool is provided on YouTube\footnote{\url{https://www.youtube.com/watch?v=pSdll8hOFjY}}. The tool is built to be freely used by anyone who aims to improve their APIs' design quality. The \textsf{REST-Ling} tool might be of interest to academics aiming to perform further research on API quality. Practitioners can also use the tool to assess the quality of their APIs.
\section{Experiments and Results}\label{sec:experiments}

This section reports on two empirical studies using the SARAv2 approach. The first study in Section \ref{sec:Overview} aims at performing the qualitative analysis of IoT APIs utilising the \textsf{REST-Ling} tool. In the second study, in Section \ref{sec:effectiveness}, we assess the effectiveness of the \textsf{REST-Ling} tool by validating the accuracy of the detection heuristics and the efficiency of the detection algorithms that are part of the underlying SOFA framework~\cite{10.1007/978-3-642-34321-61}. In the following sections, we provide the details of the study results.

\subsection{Subjects and Objects}\label{sec:subjects and objects}

In these two empirical studies, we consider nine linguistic antipatterns and their corresponding patterns as discussed in Section \ref{sec:LAPs}. As our objects, we collected a list of more than 700 Web APIs from \url{programmableweb.com} of 73 types including `Big Data', `Cloud', `Database-as-a-Service', `Infrastructure-as-a-Service', `Internet of Things', and `Platform-as-a-Service'. From that list, we filtered only APIs related to the `Internet of Things' and finally chose 19 IoT APIs that have well-organised API documentation. We manually extracted the URIs, their documentation, and underlying HTTP methods. We collected and analysed a set of 1,102 URIs from the 19 IoT APIs. Table \ref{List of APIs} lists the 19 IoT APIs and their online documentation that we analysed. We then apply detection heuristics of nine patterns and antipatterns as defined in Section \ref{sec:LAPs} on the URIs to perform syntactic and semantic analyses. For all detection, we rely on the SOFA framework~\cite{10.1007/978-3-642-34321-61}.

We inspected the documentation for the APIs to assess the support for MQTT. We found that 11 of them do not support MQTT in any way. We also found that in five of those that explicitly mention MQTT, only a subset of the full API functionality is supported, or the documentation is lacking. This leaves three APIs that claim full support for the MQTT protocol. This supports our claim that REST is the dominating style for developing cloud-based IoT applications.

\begin{table}[t!]
\caption{List of 19 analysed IoT APIs and their online documentations.}\label{List of APIs}
\setlength\tabcolsep{5pt}
\scriptsize
\begin{center}
\begin{tabular}{l c}
\toprule
\textbf{IoT APIs and Online Documentation} & \textbf{\#URIs Tested}\\
\midrule
 \href{docs.aws.amazon.com/iot/index.html}{Amazon AWS IoT Core} & 150 \\
 \href{ambrosus.docs.apiary.io/\#}{Ambrosus Gateway} & 14 \\
  \href{https://www.arduino.cc/reference/en/iot/api/}{Arduino IoT Cloud API} & 20 \\
 \href{caret.co/caret-api/}{Caret} & 7 \\
  \href{https://developer.cisco.com/docs/flare-guide/\#!flare-api}{Cisco Flare} & 34 \\
  \href{cisco.com/c/en/us/td/docs/interoperability\_systems/c\_ipics/461/api/guide/api461/restfunc.html}{Cisco IPICS} & 5\\
  \href{docs.clearblade.com/v/2/static/restapi/index.html}{ClearBlade} & 84\\
  \href{my.cubesensors.com/docs}{CubeSensors} & 4 \\
  \href{docs.droplit.io}{Droplit.io} & 52\\
  \href{developers.nest.com/reference/api-overview}{Google Nest} & 47\\
  \href{docs.internetofthings.ibmcloud.com/apis/swagger/index.html}{IBM Watson IoT} & 139\\
  \href{docs.losant.com/rest-api/overview/}{Losant} & 63 \\
  \href{docs.microsoft.com/en-us/rest/api/iothub/}{Microsoft Azure} & 210\\
 \href{https://nodered.org/docs/api/}{Node-RED} & 17 \\
 \href{developer.artik.cloud/documentation/api-reference/rest-api.html}{Samsung ARTIK} & 137\\
 \href{developer.sonos.com/reference/}{Sonos} & 49\\
 \href{www.thethingsnetwork.org/docs/applications/manager/api.html}{The Things Network} & 11\\
 \href{developers.thethings.io/v2.0/reference}{thethings.iO} & 33\\
 \href{developer.toon.eu/toonapi/apis}{Toon} &	26\\ \midrule
Total  & 1,102\\
\bottomrule
\end{tabular}
\end{center}
\end{table}

\subsection{Qualitative Analysis of IoT APIs}\label{sec:Overview}

\begin{figure}[t!]
\begin{center}
\includegraphics[width=.9\textwidth]{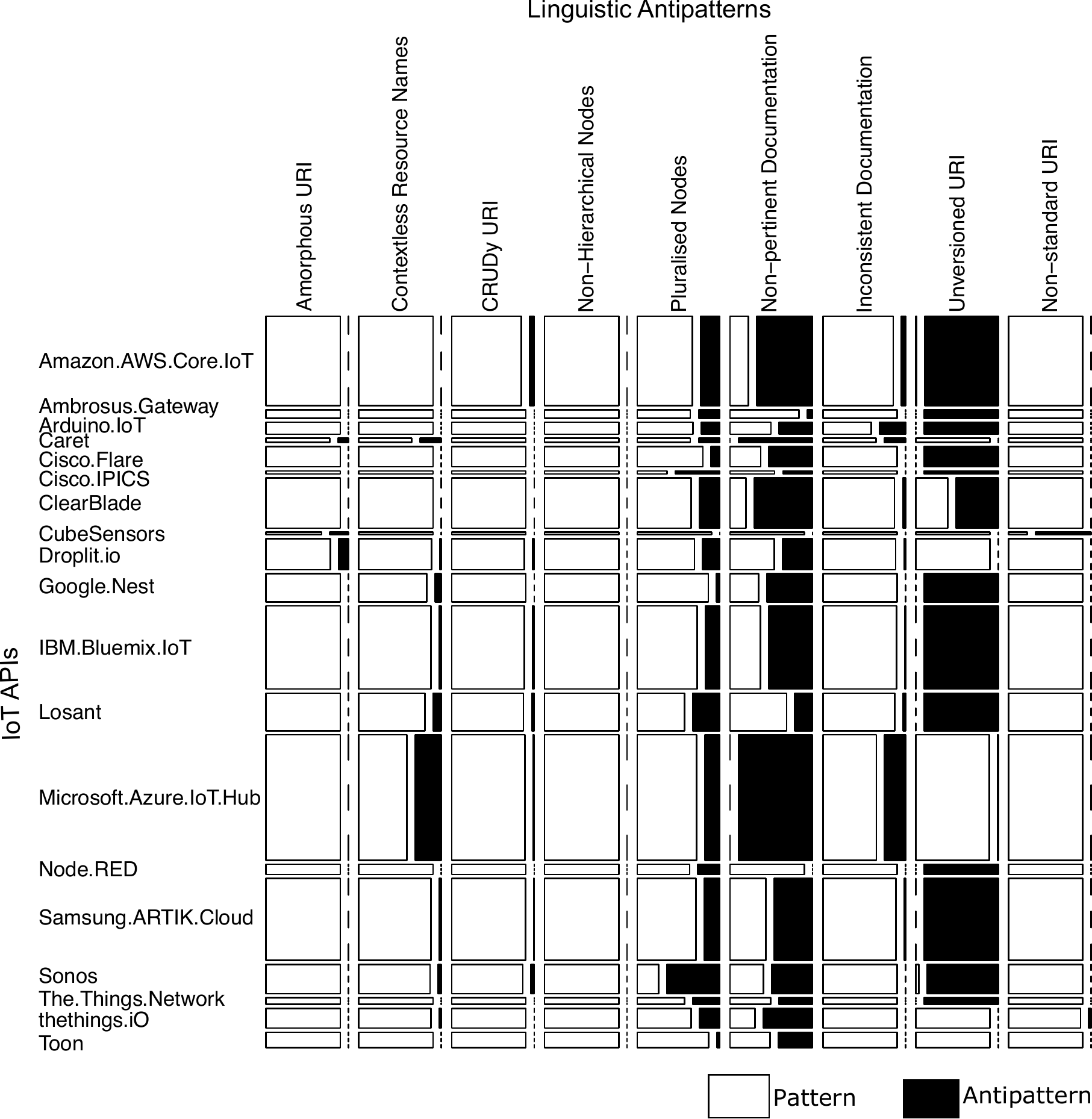}
\caption{Nine linguistic patterns and antipatterns detected in 19 IoT APIs using the \textsf{REST-Ling} tool. Columns represent patterns and antipatterns, rows represent IoT APIs with the heights of the mosaics correspond to the count of URIs analysed for each API.} \label{Detection:LA}
\end{center}
\end{figure}

This section provides an overview of the detection results. It should be noted that the detection is performed on the IoT APIs that had well-organised and self-contained documentation. Figure \ref{Detection:LA} shows the detection summary of nine linguistic patterns and antipatterns on 19 IoT APIs. In Figure \ref{Detection:LA}, columns represent patterns and antipatterns, rows represent IoT APIs with the heights of the mosaics correspond to the count of URIs analysed for each API and the colors of the mosaic correspond to \textit{white} as \textit{pattern} detection and \textit{black} as \textit{antipattern} detection.

In Figure \ref{Detection:LA}, the most frequent linguistic patterns are: (a) Tidy URI, (b) Verbless URI, (c) Hierarchical Resource Names, and (d) Standard URI. More precisely, (a) almost all of the APIs (\textit{i.e.}, 16 out of 19 analysed IoT APIs) had well-designed URIs in terms of lexical quality; (b) the majority of the analysed IoT APIs (\textit{i.e.}, 12 out of 19 analysed IoT APIs) did not include any CRUDy (Create, Read, Update, Delete, and any of their synonyms) terms or the nodes in their URIs; (c) URI nodes are well-structured in a hierarchical fashion; and (d) APIs do not tend to include special and non-English characters in their URI design. In contrast, the most frequent antipatterns are: (i) Pluralised Nodes, (ii) Non-pertinent Documentation, and (iii) Unversioned URI. In particular, (i) for the PUT/DELETE requests, the last node of the URI should be singular, and for the POST requests, the last node should be plural, however, this was not always the case for IoT APIs; (ii) the documentation was not properly aligned with the URIs; and (iii) most of the IoT API providers did not use version information within URIs, which may hinder APIs maintainability. These conclusions are based on the detection results obtained using the \textsf{REST-Ling} tool.

\begin{center}
\begin{table}[t!]
\caption{Detection results of the nine linguistic patterns and antipatterns in 19 IoT APIs.}\label{Table:Results}
\centering
\def\arraystretch{1.5}
\setlength\tabcolsep{2pt}
\fontsize{7.5}{7.5}\selectfont
\begin{tabular}{l|cc|cc|cc|cc|cc|cc|cc|cc|cc}
    \toprule
    IoT APIs & \rotatebox{90}{\textcolor{red}{\begin{scriptsize}\ding{54}\end{scriptsize}}Amorphous URI} & \rotatebox{90}{\textcolor{teal}{\begin{scriptsize}\ding{52}\end{scriptsize}}Tidy URI} & 
    \rotatebox{90}{\textcolor{red}{\begin{scriptsize}\ding{54}\end{scriptsize}}Contextless Resource Names} & \rotatebox{90}{\textcolor{teal}{\begin{scriptsize}\ding{52}\end{scriptsize}}Contextualised Resource Names} & 
    	\rotatebox{90}{\textcolor{red}{\begin{scriptsize}\ding{54}\end{scriptsize}}CRUDy URI} & 
    	\rotatebox{90}{\textcolor{teal}{\begin{scriptsize}\ding{52}\end{scriptsize}}Verbless URI} & 
    	\rotatebox{90}{\textcolor{red}{\begin{scriptsize}\ding{54}\end{scriptsize}}Non-Hierarchical Nodes} & 
    	\rotatebox{90}{\textcolor{teal}{\begin{scriptsize}\ding{52}\end{scriptsize}}Hierarchical Nodes} & 
    	\rotatebox{90}{\textcolor{red}{\begin{scriptsize}\ding{54}\end{scriptsize}}Pluralised Nodes} & 
    	\rotatebox{90}{\textcolor{teal}{\begin{scriptsize}\ding{52}\end{scriptsize}}Singularised Nodes} & 
    	\rotatebox{90}{\textcolor{red}{\begin{scriptsize}\ding{54}\end{scriptsize}}Non-pertinent Documentation} & 
    	\rotatebox{90}{\textcolor{teal}{\begin{scriptsize}\ding{52}\end{scriptsize}}Pertinent Documentation} & 
    	\rotatebox{90}{\textcolor{red}{\begin{scriptsize}\ding{54}\end{scriptsize}}Inconsistent Documentation} & 
    	\rotatebox{90}{\textcolor{teal}{\begin{scriptsize}\ding{52}\end{scriptsize}}Consistent Documentation} & 
    	\rotatebox{90}{\textcolor{red}{\begin{scriptsize}\ding{54}\end{scriptsize}}Unversioned URI} & 
    	\rotatebox{90}{\textcolor{teal}{\begin{scriptsize}\ding{52}\end{scriptsize}}URI Versioning} & 
    	\rotatebox{90}{\textcolor{red}{\begin{scriptsize}\ding{54}\end{scriptsize}}Non-standard URI} & 
    	\rotatebox{90}{\textcolor{teal}{\begin{scriptsize}\ding{52}\end{scriptsize}}Standard URI} \\
    	\midrule
Amazon AWS Core IoT & 0 & 150 & 0 & 150 & 9 & 141 & 0 & 150 & 39 & 111 & 113 & 37 & 8 & 142 & 148 & 2 & 0 & 150 \\
Ambrosus Gateway & 0 & 14 & 0 & 14 & 0 & 14 & 0 & 14 & 4 & 10 & 1 & 13 & 0 & 14 & 14 & 0 & 0 & 14 \\
Arduino IoT & 0 & 20 & 0 & 20 & 0 & 20 & 0 & 20 & 5 & 15 & 9 & 11 & 7 & 13 & 20 & 0 & 0 & 20 \\
Caret & 1 & 6 & 2 & 5 & 0 & 7 & 0 & 7 & 2 & 5 & 7 & 0 & 2 & 5 & 0 & 7 & 0 & 7 \\
Cisco Flare & 0 & 34 & 0 & 34 & 0 & 34 & 0 & 34 & 4 & 30 & 20 & 14 & 0 & 34 & 34 & 0 & 0 & 34 \\
Cisco IPICS & 0 & 5 & 0 & 5 & 0 & 5 & 0 & 5 & 3 & 2 & 2 & 3 & 0 & 5 & 5 & 0 & 0 & 5 \\
ClearBlade & 0 & 84 & 0 & 84 & 0 & 84 & 0 & 84 & 23 & 61 & 66 & 18 & 3 & 81 & 48 & 36 & 0 & 84 \\
CubeSensors & 1 & 3 & 0 & 4 & 0 & 4 & 0 & 4 & 0 & 4 & 0 & 4 & 0 & 4 & 0 & 4 & 3 & 1 \\
Droplit.io & 7 & 45 & 1 & 51 & 1 & 51 & 0 & 52 & 12 & 40 & 21 & 31 & 1 & 51 & 0 & 52 & 0 & 52 \\
Google Nest & 0 & 47 & 4 & 43 & 0 & 47 & 0 & 47 & 2 & 45 & 29 & 18 & 0 & 47 & 47 & 0 & 0 & 47 \\
IBM Bluemix IoT & 0 & 139 & 4 & 135 & 3 & 136 & 0 & 139 & 27 & 112 & 82 & 57 & 1 & 138 & 139 & 0 & 0 & 139 \\
Losant & 0 & 63 & 7 & 56 & 2 & 61 & 0 & 63 & 23 & 40 & 15 & 48 & 2 & 61 & 63 & 0 & 0 & 63 \\
Microsoft Azure IoT Hub & 0 & 210 & 74 & 136 & 3 & 207 & 0 & 210 & 42 & 168 & 210 & 0 & 59 & 151 & 2 & 208 & 0 & 210 \\
Node-RED & 0 & 17 & 0 & 17 & 0 & 17 & 0 & 17 & 5 & 12 & 0 & 17 & 0 & 17 & 17 & 0 & 0 & 17 \\
Samsung ARTIK Cloud & 0 & 137 & 4 & 133 & 1 & 136 & 0 & 137 & 29 & 108 & 71 & 66 & 2 & 135 & 137 & 0 & 0 & 137 \\
Sonos & 0 & 49 & 2 & 47 & 2 & 47 & 0 & 49 & 35 & 14 & 27 & 22 & 0 & 49 & 47 & 2 & 0 & 49 \\
The Things Network & 0 & 11 & 0 & 11 & 0 & 11 & 0 & 11 & 4 & 7 & 5 & 6 & 0 & 11 & 11 & 0 & 0 & 11 \\
thethings.iO & 0 & 33 & 1 & 32 & 0 & 33 & 0 & 33 & 9 & 24 & 22 & 11 & 0 & 33 & 0 & 33 & 1 & 32 \\
Toon & 0 & 26 & 0 & 26 & 0 & 26 & 0 & 26 & 1 & 25 & 12 & 14 & 0 & 26 & 0 & 26 & 0 & 26 \\
\bottomrule
\end{tabular}
\end{table}
\end{center}

Below, we briefly discuss the detection of some of the most and least common linguistic antipatterns.

\vspace{3mm}
\noindent\textbf{CRUDy URI:} The URI \url{/v0/api/auth/shortcode/create} from Droplit.io was detected as \textcolor{red}{\begin{scriptsize}\ding{54}\end{scriptsize}}\textit{CRUDy URI}. The POST method was used to do that. Even without adding the `create' node at the end of the URI, using the POST method, it was already understood that the goal was to create a shortcode that will be used for authentication, as stated in its documentation. The URL \url{/bulk/devices/remove} in IBM Watson IoT had a similar issue where it used the POST method to delete multiple devices. However, these poor practices of introducing CRUDy terms (or their synonyms) are highly discouraged in REST since there are a number of action-oriented HTTP methods available. The URI designers will simply combine an appropriate HTTP method from those with their URIs and perform diverse resource- or things-oriented tasks. The Samsung ARTIK also had similar issue, \textit{e.g.}, a URI \url{/trials/sessions/search} is found with the `search' node at the end.

\vspace{3mm}
\noindent\textbf{Inconsistent Documentation:} The \textcolor{red}{\begin{scriptsize}\ding{54}\end{scriptsize}}\textit{Inconsistent Documentation} refers to the case where the HTTP method applies with a URI that has an opposite documentation, \textit{i.e.}, the HTTP method does not do what it says. This is similar to \textit{Method Signature and Comment are Opposite} linguistic antipattern in object-oriented programming \cite{Arnaoudova2016, Arnaoudova2013}. For example, from the Droplit.io API, there is a URI \url{/v0/api/clients/} that was applied with a GET method, but has the documentation as `\textit{Create a client. An account token or server token may...}'. Clearly, using a GET method to create a client is poor (or even wrong) practice in REST. Similar instances were found in IBM Watson IoT, \textit{e.g.}, it uses the POST method with the URI \url{/bulk/devices/remove} and has the documentation as `\textit{delete multiple devices, each request can contain a...}'.

\vspace{3mm}
\noindent\textbf{Contextless Resources Names:} One URI \url{/devices/thermostats/device_id/time_to_target_training} by Google Nest was detected as \textcolor{red}{\begin{scriptsize}\ding{54}\end{scriptsize}}\textit{Contextless Resources Names}. We suspect that the URI was detected as an antipattern because the nodes \{\texttt{devices, thermostats, time, target, training}\} seem not to be related from the semantic point of view. When we built the topic model for Google Nest, we found that `device' and `thermostat' were present in our topic model (under the first topic cluster) and the other three node words `time', `target', and `training' were not in the topic model at all, \textit{i.e.}, they were not so important in the context of Google Nest, thus, they were considered irrelevant to the context. In our topic model, similar words or keywords that are highly related are grouped under the same topic cluster. Therefore, the \textsf{REST-Ling} tool identified the URI as \textit{Contextless Resources Names} antipattern.

\vspace{3mm}
\noindent\textbf{Non-pertinent Documentation:} Among the 19 analysed IoT APIs, all the APIs except two (\textit{i.e.}, CubeSensors and Node-RED) have instances of this antipattern. Thus, \textit{Non-pertinent Documentation} is found to be the most common antipattern. Also, 65\% of the analysed URIs (\textit{i.e.}, 712 out of 1,102) are involved in this antipattern. These findings suggest that majority of the APIs (\textit{i.e.}, 17 out of 19 analysed IoT APIs) do not provide documentation cohesive to their URI design. Also, large vendors like Amazon AWS Core IoT, IBM Watson IoT, and Google Nest do not tend to provide high-quality documentation for their APIs. For example, 75\% of the analysed URIs from Amazon AWS Core IoT, 59\% of the analysed URIs from IBM Watson IoT, and 62\% of the analysed URIs from Google Nest had \textit{Non-pertinent Documentation} antipattern. However, Microsoft Azure had 100\% of the analysed URIs provided good quality documentation, i.e., the URIs and their documentation are cohesive.

\vspace{3mm}
\noindent\textbf{Unversioned URI:} We also found a similar prevalence for \textit{Unversioned URI} antipattern, i.e., 14 out of 19 analysed IoT APIs do not include version information as part of their URIs design. In the literature, as part of the best practices, practitioners suggested including version information within the URIs \cite{RESTAPIVersioning}. This is because APIs evolve continuously, and, if not properly versioned, clients might break. In other words, versioned APIs facilitate easy maintenance both for API providers and client developers. However, it could be due to that APIs for IoT applications do not evolve frequently. Notably, some APIs are found to be in two different modes, \textit{i.e.}, some URIs are versioned whereas others are not. This observation ought not to be generalised without further investigation. For example, Amazon AWS Core IoT is found to have both unversioned and versioned URIs (148 \textit{vs.} 2) out of 150 analysed URIs. Same for Microsoft Azure, where we found 2 out of 210 analysed URIs had version info included in the URI design. Thus, there is a clear lack of standardised practice among the API providers.

\vspace{3mm}
\noindent\textbf{Non-Standard URI:} According to the definition of \textit{Non-standard URI} antipattern in Section \ref{nonstandardURI}, URI design should not include nodes or resources with non-standard identification (\textit{e.g.}, special or unknown characters, blank spaces, double hyphens, etc.), which hinders the reusability and understandability of the APIs. Our findings suggest that the majority of the IoT APIs, \textit{i.e.}, 17 out of 19 analysed APIs, follow the standard URI design practices. From the CubeSensors API, three URIs are found: \url{/devices/[device id]}, \url{/devices/[device id]/current}, and \url{/devices/[device id]/span} that a had blank space as part of the URI design. Also, the thethings.iO API had this URI \url{/things/THING_TOKEN/resources/$MAGIC_RESOURCE} with a dollar sign (\$) before the parameter, which is considered as an unknown character in the URI design. These practices of URI design make the URI non-standard and hinder the reusability and understandability of the APIs.


\subsection{Effectiveness of the \textsf{REST-Ling}}\label{sec:effectiveness}
This section answers our research questions in showing the effectiveness of the \textsf{REST-Ling} tool.

\subsubsection{Research Questions}\label{sec:Hypotheses}

We define four research questions related to the prevalence, accuracy, and to assess the usefulness and effectiveness of the \textsf{REST-Ling} tool that is developed based on the SARAv2 approach.

\begin{itemize}
    \item RQ$_1$ Prevalence: \textit{To what extent IoT APIs suffer from poor linguistic design quality, i.e., linguistic antipatterns?} With RQ$_1$, we want to investigate whether IoT APIs suffer from linguistic design quality (\textit{i.e.}, linguistic antipatterns), and to what extent.
    
    \item RQ$_2$ Comparison: \textit{To what extent APIs across domains suffer from poor linguistic design quality, i.e., linguistic antipatterns?} With RQ$_2$, we want to investigate whether and to what extent APIs for Web applications and cloud services are  prone to linguistic antipatterns than the APIs for IoT applications.
    
    \item RQ$_3$ Accuracy: \textit{What is the accuracy of \textsf{REST-Ling} on the detection of linguistic antipatterns?} With RQ$_3$, we want to investigate the accuracy of our defined detection heuristics implemented in the \textsf{REST-Ling} tool.
    
        \item RQ$_4$ Efficiency: \textit{How does the \textsf{REST-Ling} perform in terms of average detection time for linguistic antipatterns?} With RQ$_4$, we want to study the detection performance of the \textsf{REST-Ling} tool in executing the detection heuristics implemented as part of the tool. We conjecture that an average detection time in the order of seconds is acceptable for each antipattern.
    
\end{itemize}

\subsubsection{Validation Process}\label{sec:validation process}

We applied random sampling to choose 91 URIs out of 1,102 URIs from 19 APIs to measure the overall accuracy of the \textsf{REST-Ling} tool. The total population size is 9,918 (\textit{i.e.}, 1,102 URIs $\times$ 9 patterns). We aim for a 95\% confidence level and a confidence interval of 10; thus, a sample size of 819 questions (\textit{i.e.}, 91 URIs $\times$ 9 antipatterns) were selected to be validated manually. We involved three professionals to manually validate our detection findings. The professionals have knowledge on REST and were not part of the implementation and execution of the detection algorithms. Two of the professionals co-author this paper and have both industrial and academic experience, and the third is an industry expert with working knowledge on Web APIs. To avoid any potential conflicts of interest and to be fully transparent, during the experiments and analyses, the obtained detection results were not shared and discussed with any of the authors who later participated in the validation process. In this way, we ensured that the accuracy of the detection results was not affected, \textit{i.e.}, the accuracy measurements were unbiased.

To facilitate the validation process, the textual descriptions of linguistics patterns and antipatterns, the URIs along with the HTTP method, and their documentations were provided. To set the oracle, we decided on the majority. That is, each detection instance is manually validated by three participants and the oracle is decided when at least two participants accept or reject an instance. The validation process with questions and responses is done online and available using Google Forms\footnote{\url{https://forms.gle/EmGPoZbRwGXybFHY8}}. As shown in Equation 1, the \textit{accuracy} measure was used to measure the performance of the \textsf{REST-Ling} tool. We also use the Matthews' correlation coefficient (MCC), as shown in Equation 2, as an alternative measure unaffected by the unbalanced datasets issue. The MCC is computed based on the contingency matrix. MCC generates a higher score if our \textsf{REST-Ling} tool as a binary classifier can correctly classify the majority of antipattern instances and the majority of the pattern instances. MCC has the values –1 and +1 for a perfect misclassification and perfect classification, respectively.

\begin{small}
\begin{equation}
    Accuracy = \frac{ TP + TN }{ TP + TN + FP + FN}
\end{equation}
\begin{equation}
    MCC = \frac{ TP \times TN - FP \times FN }{ \sqrt{(TP + FP)(TP + FN)(TN + FP)(TN + FN)} }
\end{equation}
\noindent TP: True Positive; FP: False Positive; TN: True Negative; and FN: False Negative.
\end{small}

\vspace{2mm}
In the following sections, we answer the four research questions as stated in Section \ref{sec:Hypotheses}.

\subsubsection{RQ$_1$ Prevalence}

Table \ref{Table:Results} presents detection results for the nine pairs of linguistic patterns and antipatterns on 19 IoT APIs. In Table \ref{Table:Results}, the first column shows the linguistic patterns and antipatterns followed by the 19 IoT APIs. For each API and for each pattern and antipattern, the total number of occurrences is reported that are found as positives by our detection algorithms. The last column shows the total occurrences with percentage for each pattern and antipattern. The detailed analyses results for all the 1,102 URIs from 19 IoT APIs are available online\footnote{\url{https://doi.org/10.5281/zenodo.6393404}}.

To summarise the results in Table \ref{Table:Results}, Amazon AWS Core IoT, Arduino IoT, Caret, Cisco IPICS, and Sonos had the most number of antipatterns given the number of URIs tested for each of those APIs. In contrast, Google Nest, Node-RED, CubeSensors, thethings.iO, and Toon had the most number of patterns given the number of URIs tested for each those APIs. We make these observations by dividing the total instances of patterns or antipatterns by the total number of analysed URIs for each API.

As Table \ref{Table:Results} suggests, all the analysed IoT APIs contain at least one of nine linguistic antipatterns. The set of five IoT APIs, \textit{i.e.}, Droplit.io, IBM Watson IoT, Losant, Microsoft Azure IoT Hub, and Samsung ARTIK Cloud, is found to be involved in six different linguistic antipatterns. Although, CubeSensors, Node-RED, and Toon APIs are found to be involved in only two linguistic antipatterns.

\begin{center}
\begin{tcolorbox}[colback=gray!15,
                  colframe=gray,
                  width=11cm,
                  left=1pt,
                  right=1pt,
                  top=1pt,
                  bottom=1pt,
                  arc=1mm, auto outer arc]
\begin{small}
 \textbf{Summary on RQ$_1$:} \textbf{Linguistic antipatterns are prevalent in IoT APIs.} In the analysed IoT APIs we have detected some instances of poor design practices, being the most prevalent \textit{Non-pertinent Documentation} or \textit{Unversioned URIs}. We also observed the presence of good design practices, \textit{i.e.}, linguistic patterns, which suggests that the developers are aware of the need for linguistic quality on their APIs.
\end{small} 
\end{tcolorbox}
\end{center}

\subsubsection{RQ$_2$ Comparison}
  
\begin{table}[t!]
\caption{Comparison of the detection of linguistic antipatterns across domains.}\label{tab:ComparisonDomains}
    \centering
    \scriptsize 
    \def\arraystretch{1}
\setlength\tabcolsep{1.75pt}
    \begin{tabular}{lcrcrcr}
    \toprule
Approach & \multicolumn{2}{c}{\textbf{SARA} \cite{Palma2017SemanticAO}} & \multicolumn{2}{c}{\textbf{CloudLex} \cite{10.1007/978-3-319-94959-816}} & 	\multicolumn{2}{c}{\textbf{SARAv2}} \\ \midrule
Target domain & \multicolumn{2}{c}{APIs for Web apps}	 & \multicolumn{2}{c}{APIs for Cloud services}	 & \multicolumn{2}{c}{APIs for IoT}	 \\
Number of APIs Analysed & \multicolumn{2}{c}{18}	 & \multicolumn{2}{c}{16} & 	\multicolumn{2}{c}{19}	 \\
Number of URIs Tested & \multicolumn{2}{c}{310}	 & \multicolumn{2}{c}{23,062} & 	\multicolumn{2}{c}{1,102}	 \\ \midrule
(Anti)Patterns / Detection & \#Instances & \%URIs & \#Instances & \%URIs & \#Instances & \%URIs \\ \midrule
\textcolor{red}{\begin{scriptsize}\ding{54}\end{scriptsize}}Amorphous URI & 202 & 65\% & - & - & 9 & 0.82\% \\
\textcolor{teal}{\begin{scriptsize}\ding{52}\end{scriptsize}}Tidy URI & 108 & 35\% & - & - & 1,093 & 99.18\% \\
\textcolor{red}{\begin{scriptsize}\ding{54}\end{scriptsize}}Contextless Resource Names & 123 & 40\% & 10,595 & 23\% & 99 & 8.98\% \\
\textcolor{teal}{\begin{scriptsize}\ding{52}\end{scriptsize}}Contextualised Resource Names & 187 & 60\% & 12,467 & 77\% & 1,003 & 91.02\% \\
\textcolor{red}{\begin{scriptsize}\ding{54}\end{scriptsize}}CRUDy URI & 38 & 12\% & - & - & 21 & 1.91\% \\
\textcolor{teal}{\begin{scriptsize}\ding{52}\end{scriptsize}}Verbless URI & 272 & 88\% & - & - & 1,081 & 98.09\% \\
\textcolor{red}{\begin{scriptsize}\ding{54}\end{scriptsize}}Inconsistent Documentation & - & - & - & - & 85 & 7.71\% \\
\textcolor{teal}{\begin{scriptsize}\ding{52}\end{scriptsize}}Consistent Documentation & - & - & - & - & 1,017 & 92.29\% \\
\textcolor{red}{\begin{scriptsize}\ding{54}\end{scriptsize}}Non-hierarchical Nodes & 173 & 56\% & - & - & 0 & 0\% \\
\textcolor{teal}{\begin{scriptsize}\ding{52}\end{scriptsize}}Hierarchical Nodes & 0 & 0\% & - & - & 1,102 & 100\% \\
\textcolor{red}{\begin{scriptsize}\ding{54}\end{scriptsize}}Non-pertinent Documentation & 155* & 28\% & 1,339** & 52\% & 712 & 64.61\% \\
\textcolor{teal}{\begin{scriptsize}\ding{52}\end{scriptsize}}Pertinent Documentation & 400* & 72\% & 792** & 48\% & 390 & 35.39\% \\
\textcolor{red}{\begin{scriptsize}\ding{54}\end{scriptsize}}Pluralised Nodes & 14 & 5\% & - & - & 269 & 24.41\% \\
\textcolor{teal}{\begin{scriptsize}\ding{52}\end{scriptsize}}Singularised Nodes & 4 & 1\% & - & - & 833 & 75.59\% \\
\textcolor{red}{\begin{scriptsize}\ding{54}\end{scriptsize}}Unversioned URI & - & - & - & - & 732 & 66.42\% \\
\textcolor{teal}{\begin{scriptsize}\ding{52}\end{scriptsize}}Versioned URI & - & - & - & - & 370 & 33.58\% \\
\textcolor{red}{\begin{scriptsize}\ding{54}\end{scriptsize}}Non-standard URI & - & - & - & - & 4 & 0.36\% \\
\textcolor{teal}{\begin{scriptsize}\ding{52}\end{scriptsize}}Standard URI & - & - & - & - & 1,098 & 99.64\% \\
\bottomrule
\multicolumn{7}{l}{\textit{*Detection was done on additional set of URIs  **detection was done on a subset of the URIs}}\\
    \end{tabular}
\end{table}

APIs are used for various purposes and in various domains, for example, APIs for Web applications~\cite{Palma2017SemanticAO} and Cloud services~\cite{10.1007/978-3-319-94959-816}. In this research, we aim to find whether a certain linguistic antipattern (or pattern) is notably common across the domains or whether a certain domain is more prone to an antipattern (or pattern) compared to other domains. 
It is important to note that the methods relevant to the domains, i.e., SARA \cite{Palma2017SemanticAO} and CloudLex \cite{10.1007/978-3-319-94959-816} are applied to different sets of APIs. Thus, a direct comparison among the methods is not possible. Instead, we only want to compare the prevalence of linguistic antipatterns in REST APIs in different domains.


As Table \ref{tab:ComparisonDomains} shows, the APIs for Web applications have more \textcolor{red}{\begin{scriptsize}\ding{54}\end{scriptsize}}\textit{Amorphous URI} (65\%) than the IoT APIs (0.82\%). In contrast, IoT APIs are more often tidy than the APIs from the other domains with 99.18\% URIs are detected as \textcolor{teal}{\begin{scriptsize}\ding{52}\end{scriptsize}}Tidy URI. A consistent detection is observed in Table \ref{tab:ComparisonDomains} for the \textcolor{red}{\begin{scriptsize}\ding{54}\end{scriptsize}}\textit{Contextless Resource Names} and its corresponding \textcolor{teal}{\begin{scriptsize}\ding{52}\end{scriptsize}}\textit{Contextualised Resource Names}, \textit{i.e.}, the majority of the URIs in the three domains are designed with resource names that are semantically aligned within the context of the URI design. Thus, 60\% of URIs for Web applications, 77\% of URIs for Cloud services, and 91.02\% of URIs for IoT APIs are detected as \textcolor{teal}{\begin{scriptsize}\ding{52}\end{scriptsize}}\textit{Contextualised Resource Names}. Similarly, designers are well aware of not using verbs within the URI design, thus, 88\% URIs for Web applications and 98.09\% URIs from the IoT APIs are detected as \textcolor{teal}{\begin{scriptsize}\ding{52}\end{scriptsize}}\textit{Verbless URI}. To summarise, APIs for Web applications are mostly prone to \textcolor{red}{\begin{scriptsize}\ding{54}\end{scriptsize}}\textit{Amorphous URI} and notably implement patterns like \textcolor{teal}{\begin{scriptsize}\ding{52}\end{scriptsize}}\textit{Verbless URI} and \textcolor{teal}{\begin{scriptsize}\ding{52}\end{scriptsize}}\textit{Pertinent Documentation}.  On the other hand, IoT APIs suffer mostly with \textcolor{red}{\begin{scriptsize}\ding{54}\end{scriptsize}}\textit{Non-pertinent Documentation} and \textcolor{red}{\begin{scriptsize}\ding{54}\end{scriptsize}}\textit{Unversioned URI}, which suggests that IoT APIs are poorly documented and the URI designers do not tend to design the URIs with version info -- a poor URI design practice. In contrast, we found that IoT APIs have very tidy URIs, \textit{i.e.}, the \textcolor{teal}{\begin{scriptsize}\ding{52}\end{scriptsize}}\textit{Tidy URI}, and the nodes in the URIs are organised hierarchically, \textit{i.e.}, the \textcolor{teal}{\begin{scriptsize}\ding{52}\end{scriptsize}}\textit{Hierarchical Nodes}.

More specifically, for Web APIs in~\cite{Palma2017SemanticAO}, for example, Facebook had a high number of \textcolor{red}{\begin{scriptsize}\ding{54}\end{scriptsize}}\textit{Contextless Resource Names} and \textcolor{red}{\begin{scriptsize}\ding{54}\end{scriptsize}}\textit{Non-hierarchical Nodes} due to its diverse and large set of resources. It is often difficult to find a best hierarchical order of URIs nodes or to find resources names best fit to a certain context. However, Twitter and YouTube, for example, did not suffer those antipatterns with comparatively lower number of resources than Facebook \cite{Palma2017SemanticAO}. In fact, on average, StackExchange had the most number of antipatterns, due to which \textcolor{red}{\begin{scriptsize}\ding{54}\end{scriptsize}}\textit{Amorphous URI} and \textcolor{red}{\begin{scriptsize}\ding{54}\end{scriptsize}}\textit{Non-hierarchical Nodes} seem very common, as reported by SARA~\cite{Palma2017SemanticAO}.

On the contrary, relatively new IoT APIs are designed with more knowledge and experience from the literature of good design practices and guidelines on APIs design \cite{Masse2012, RESTfulBestPractices, MicrosoftMSDN, RESTAPIVersioning, IoTDesign1, IoTDesign2}. This could be one reason the \textcolor{red}{\begin{scriptsize}\ding{54}\end{scriptsize}}\textit{Amorphous URI} is found on very small scale. Also, the detection for \textcolor{red}{\begin{scriptsize}\ding{54}\end{scriptsize}}\textit{Contextless Resource Names} and \textcolor{red}{\begin{scriptsize}\ding{54}\end{scriptsize}}\textit{Non-hierarchical Nodes} in IoT APIs resulted in comparatively lower than in APIs for Web applications. The major IoT APIs vendors including Amazon, Google, IBM, and Microsoft are well aware of designing quality URIs both syntactic (\textit{e.g.}, \textcolor{red}{\begin{scriptsize}\ding{54}\end{scriptsize}}\textit{Amorphous URI}) and semantic (\textit{e.g.}, \textcolor{red}{\begin{scriptsize}\ding{54}\end{scriptsize}}\textit{Contextless Resource Names}, \textcolor{red}{\begin{scriptsize}\ding{54}\end{scriptsize}}\textit{Non-hierarchical Nodes}, or \textcolor{red}{\begin{scriptsize}\ding{54}\end{scriptsize}}\textit{Inconsistent Documentation}) viewpoints.

Overall, on average, 34\% of the URIs from the APIs for Web applications are detected having linguistic antipatterns by SARA~\cite{Palma2017SemanticAO}. In contrast, only 17\% of the URIs from the APIs for IoT devices are detected as antipatterns by the \textsf{REST-Ling} tool. Also, for linguistic patterns, the \textsf{REST-Ling} tool found 73\% URIs are well-designed compared to 42\% URIs of APIs for Web applications. This suggests that IoT APIs are comparatively more well-designed than the APIs for Web applications like Facebook, YouTube, or Instagram~\cite{Palma2017SemanticAO}. This could be because the APIs specific to Web applications deal with a plethora of resources types and representations, compared to the APIs in the IoT domain, where devices mainly deal with device data and transmission from/to the servers and peer devices. Thus, APIs for Web applications pose a higher challenge in designing high-quality URIs than the IoT APIs, \textit{i.e.}, APIs for Web applications are more prone to linguistic antipatterns.

\begin{center}
\begin{tcolorbox}[colback=gray!15,
                  colframe=gray,
                  width=11cm,
                  left=1pt,
                  right=1pt,
                  top=1pt,
                  bottom=1pt,
                  arc=1mm, auto outer arc]
\begin{small}  \textbf{Summary on RQ$_2$:} We found that resource URIs are structurally and contextually well-designed in APIs for IoT applications than for Web applications. Although the APIs for cloud services are not studied to a large number, the analysis of resource context (\textit{Contextless vs. Contextualised Resource Names}) and cohesive documentation (\textit{Pertinent vs. Non-pertinent Documentation}) suggests that APIs for cloud services exhibit similar design quality found in APIs for Web applications. In fact, \textbf{APIs for IoT applications appear to have a better design (structural and contextual) except that the APIs for IoT applications are poorly and, in many cases, briefly documented}.
\end{small}
\end{tcolorbox}
\end{center}

\subsubsection{RQ$_3$ Accuracy}

\begin{table}[t!]
\caption{\textsf{REST-Ling} validation results to compute overall precision and recall.}\label{Table: Validation 1}
\centering
\def\arraystretch{1.5}
\setlength\tabcolsep{2pt}
\fontsize{8}{8}\selectfont
\begin{tabular}{lccccccccc}
\toprule
\textbf{Linguistic Antipatterns} &  \textbf{P} & \textbf{N}  & \textbf{TP} & \textbf{FP}  & \textbf{FN} & \textbf{TN}  & \textbf{Accuracy} & \textbf{MCC}  \\
\toprule
\textcolor{red}{\begin{scriptsize}\ding{54}\end{scriptsize}}\textit{Amorphous URI} & 4 & 87 & 1 & 3 & 27 & 60 & 67\% & -0.03 \\
\midrule
\textcolor{red}{\begin{scriptsize}\ding{54}\end{scriptsize}}\textit{Contextless Resource Names} & 11 & 80 & 6 & 5 & 15 & 65 & 78\% & 0.28  \\
\midrule
\textcolor{red}{\begin{scriptsize}\ding{54}\end{scriptsize}}\textit{CRUDy URI} & 3 & 88 & 3 & 0 & 3 & 85 & 97\% & 0.69 \\
\midrule
\textcolor{red}{\begin{scriptsize}\ding{54}\end{scriptsize}}\textit{Non-hierarchical Nodes} & 0 & 91 & 0 & 0 & 14 & 77 & 85\% & \textit{n/a} \\
\midrule
\textcolor{red}{\begin{scriptsize}\ding{54}\end{scriptsize}}\textit{Pluralised Nodes} & 31 & 60 & 25 & 6 & 5 & 55 & 88\% & 0.73 \\
\midrule
\textcolor{red}{\begin{scriptsize}\ding{54}\end{scriptsize}}\textit{Non-Pertinent Documentation} & 58 & 33 & 8 & 50 & 4 & 29 & 41\% & 0.02 \\
\midrule
\textcolor{red}{\begin{scriptsize}\ding{54}\end{scriptsize}}\textit{Unversioned URIs} & 67 & 24 & 67 & 0 & 0 & 24 & 100\% & 1.00 \\
\midrule
\textcolor{red}{\begin{scriptsize}\ding{54}\end{scriptsize}}\textit{Inconsistent Documentation} & 26 & 65 & 19 & 7 & 8 & 57 & 84\% & 0.60 \\
\midrule
\textcolor{red}{\begin{scriptsize}\ding{54}\end{scriptsize}}\textit{Non-standard URI} & 2 & 89 & 2 & 0 & 10 & 79 & 89\% & 0.38 \\
\bottomrule
\textbf{Total} & \textbf{202} & \textbf{617} & \textbf{131} & \textbf{71} & \textbf{86} & \textbf{531} &  &  \\
\textbf{Average} &  &  &  &  &  &  & \textbf{81\%} & \textbf{0.46} \\
\bottomrule
\end{tabular}
\end{table}

Table \ref{Table: Validation 1} shows the detection accuracy for nine linguistic antipatterns. In Table \ref{Table: Validation 1}, on a subset of 91 URIs from 19 IoT APIs, we obtained an average accuracy of 81\%. The accuracy is also heavily dependent on how engineers (in our case, the three professionals) understand and interpret a phrase or word based on their experience and knowledge. For example, in Validation 1, an instance from Losant with the URI \url{/applications/APPLICATION_ID/devices/DEVICE_ID/commandStream} and the documentation\footnote{\footnotesize{\texttt{Attach to a real time stream of command messages to this device using SSE.}}} where the \textsf{REST-Ling} tool detects it as \textit{Non-pertinent Documentation} antipattern, but the majority of the professionals (\textit{i.e.}, two out of three) considered the URI and its documentation cohesive, thus, decided as \textit{Pertinent Documentation} pattern. In another example, an instance from Cisco Flare with the URI \url{/environments/{environment_id}/zones/{zone_id}/things/{thing_id}/data} and with the documentation\footnote{\footnotesize{\texttt{Get thing data. Gets all data values for a thing by making a REST call. You can also get the data for a thing using the getData Socket.IO call.}}} where the \textsf{REST-Ling} tool approach detects it as \textit{Pertinent Documentation} pattern, however, two out of three professionals did not see this URI and its documentation cohesive, and identified it as \textit{Non-pertinent Documentation} antipattern. Instances similar to the above examples may lead to lower accuracy.

\begin{center}
\begin{tcolorbox}[colback=gray!15,
                  colframe=gray,
                  width=11cm,
                  left=1pt,
                  right=1pt,
                  top=1pt,
                  bottom=1pt,
                  arc=1mm, auto outer arc]
\begin{small}  \textbf{Summary on RQ$_3$:} The manual validation suggests that the \textsf{REST-Ling} tool has an overall \textbf{average accuracy of more than 80\%}, with an average MCC of 0.46.
\end{small}
\end{tcolorbox}
\end{center}

\begin{table}[t!]
\centering
\caption{Detection times (in seconds) of the nine linguistic patterns and antipatterns in 19 IoT APIs.}\label{Table: Detection Times}
\def\arraystretch{1}
\setlength\tabcolsep{1.75pt}
\scriptsize
\newcommand*\rot{\rotatebox{90}}
\begin{tabular}{lrrrrrrrrrrr}
\toprule
\multicolumn{1}{c}{\textbf{IoT APIs}}   & \multicolumn{1}{c}{\rot{Amorphous URI}} & \multicolumn{1}{c}{\rot{CRUDy URI}}  & \multicolumn{1}{c}{\rot{Non-hierarchical Nodes}} & \multicolumn{1}{c}{\rot{Pluralised Nodes}} & \multicolumn{1}{c}{\rot{Unversioned URI}} & \multicolumn{1}{c}{\rot{Non-pertinent Doc.}} & \multicolumn{1}{c}{\rot{Contextless Resource}} & \multicolumn{1}{c}{\rot{Inconsistent Doc.}} & \multicolumn{1}{c}{\rot{Non-standard URI}} &  \multicolumn{1}{c}{\rot{\textbf{Total}}}   & \multicolumn{1}{c}{\rot{\textbf{Average}}}  \\ \midrule
Amazon AWS IoT & 0.003 & 0.033 & 1.307 & 0.009 & 0.004 & 59.462 & 193.442 & 0.010 & 0.002 &  254.272  &  28.252 \\
Amrosus Gateway & 0.002 & 0.005 & 1.356 & 0.005 & 0.002 & 2.092 & 0.777 & 0.012 & 0.001 &  4.252  &  0.472 \\
Arduino IoT & 0.002 & 0.009 & 1.278 & 0.003 & 0.002 & 2.833 & 3.103 & 0.010 & 0.001 &  7.241  &  0.805 \\
Caret & 0.001 & 0.003 & 1.297 & 0.021 & 0.001 & 1.142 & 1.544 & 0.013 & 0.001 &  4.023  &  0.447 \\
Cisco Flare & 0.004 & 0.014 & 1.292 & 0.005 & 0.004 & 10.288 & 4.405 & 0.015 & 0.001 &  16.028  &  1.781 \\
Cisco IPICS & 0.002 & 0.005 & 1.342 & 0.004 & 0.001 & 1.037 & 0.459 & 0.010 & 0.001 &  2.861  &  0.318 \\
ClearBlade & 0.003 & 0.015 & 1.306 & 0.010 & 0.006 & 11.681 & 39.152 & 0.014 & 0.001 &  52.188  &  5.799 \\
CubeSensors & 0.001 & 0.002 & 1.317 & 0.002 & 0.001 & 1.667 & 0.610 & 0.014 & 0.001 &  3.615  &  0.402 \\
Droplit.io & 0.002 & 0.009 & 1.331 & 0.008 & 0.004 & 14.200 & 14.264 & 0.011 & 0.001 &  29.830  &  3.314 \\
Google Nest & 0.047 & 0.008 & 1.166 & 0.006 & 0.005 & 19.223 & 10.057 & 0.010 & 0.001 &  30.523  &  3.391 \\
IBM Watson & 0.004 & 0.060 & 1.275 & 0.012 & 0.007 & 75.337 & 187.547 & 0.010 & 0.040 &  264.292  &  29.366 \\
Losant & 0.003 & 0.029 & 1.244 & 0.008 & 0.003 & 12.385 & 18.489 & 0.014 & 0.002 &  32.177  &  3.575 \\
Microsoft Azure & 0.011 & 0.052 & 1.302 & 0.019 & 0.041 & 109.248 & 351.087 & 0.012 & 0.006 &  461.778  &  51.309 \\
Node-RED & 0.009 & 0.003 & 1.347 & 0.004 & 0.002 & 1.703 & 1.564 & 0.013 & 0.001 &  4.646  &  0.516 \\
Samsung ART & 0.046 & 0.018 & 1.250 & 0.012 & 0.005 & 47.180 & 125.294 & 0.011 & 0.002 &  173.818  &  19.313 \\
Sonos & 0.089 & 0.010 & 1.214 & 0.009 & 0.005 & 31.766 & 21.943 & 0.012 & 0.001 &  55.049  &  6.117 \\
The Things Net & 0.016 & 0.004 & 1.227 & 0.003 & 0.001 & 2.875 & 1.400 & 0.009 & 0.001 &  5.536  &  0.615 \\
thethings.io & 0.044 & 0.011 & 1.280 & 0.004 & 0.007 & 13.009 & 7.154 & 0.010 & 0.001 &  21.520  &  2.391 \\
Toon & 0.040 & 0.088 & 1.197 & 0.005 & 0.004 & 8.006 & 2.727 & 0.010 & 0.001 &  12.078  &  1.342 \\ \midrule
\textbf{Total} & 0.329 & 0.378 & 24.328 & 0.149 & 0.105 & 425.134 & 985.018 & 0.220 & 0.066 &  1,435.727  & \\
\textbf{Average} & 0.017 & 0.020 & 1.280 & 0.008 & 0.006 & 22.375 & 51.843 & 0.012 & 0.003 &  &  \textbf{8.396} \\
\bottomrule
\end{tabular}
\end{table}

\subsubsection{RQ$_4$ Efficiency}

We performed the experiments on an Intel Dual Core at 3.30GHz with 4GB of RAM. For the detection of linguistic patterns and antipatterns in IoT APIs, the reported detection times include: (1) the time to apply and run the detection algorithms implemented in Java on the URIs and (2) the time to export the results. Listing 2 shows the code snippet we use for measuring the detection time for each API and for each linguistic antipattern where we record the time before and after running the detection (lines 4 and 6). We then take the difference and get the values in seconds (line 7). Table \ref{Table: Detection Times} shows the detection time for each API (rows) and each antipattern (columns).

\lstset{showstringspaces=false, numbersep=1pt, breaklines=true, xleftmargin=1pt, xrightmargin=1pt, keywordstyle=\color{blue},numbers=left, stepnumber=1, language=Java, caption=Code snippet for measuring the detection time.}
\begin{lstlisting}[style=interfaces, basicstyle=\scriptsize]
    /*Detection of Contextless Resource Names Antipattern*/
    long startTime = System.currentTimeMillis();
    detectContextlessResourceNames();
    long endTime = System.currentTimeMillis();
    System.out.println("Total detection time (sec): " + (double)(endTime - startTime)/1000);
\end{lstlisting}

\begin{figure}[ht!]
\begin{center}
\includegraphics[width=0.6\textwidth]{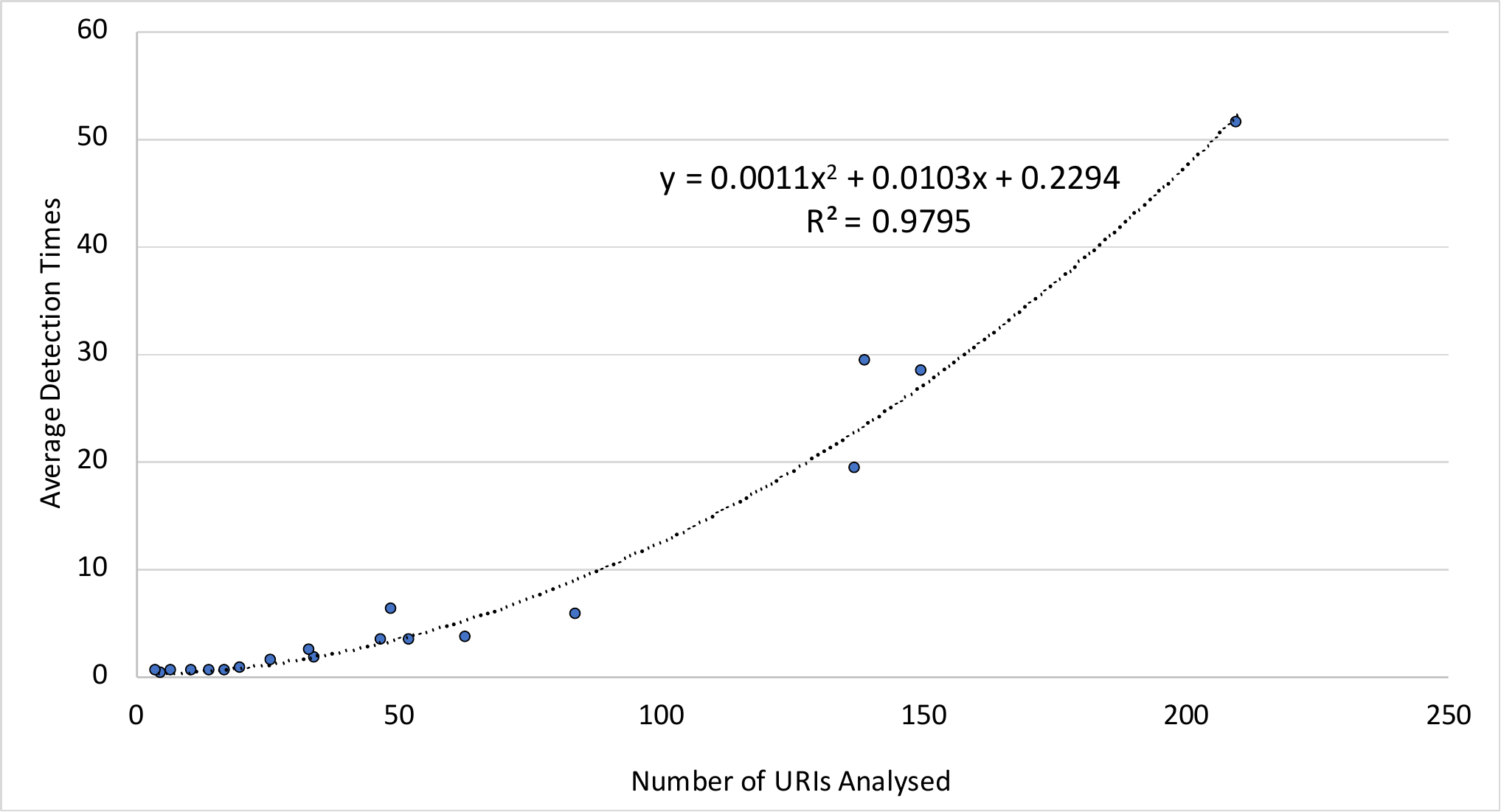}
\hfill
\caption{Plot of the average detection times against the URIs analysed for the APIs.} \label{fig:URIwiseAvgDetectTime}
\end{center}
\end{figure}

Regardless of the APIs, for each antipattern, we observed a consistent detection time, \textit{i.e.}, when the number of tested URIs for an API is low, the detection time is lower and the detection time increases polynomially when the number of tested URIs increases. The estimated growth function is $y = 0.0011x^{2}+0.0103x+0.2294$ where \textit{x} is the number of URIs to test, and \textit{y} is the total detection time for an antipattern. This growth function is particularly applicable for the \textit{Contextless Resource Names} antipattern where we needed to perform a significant amount of pairwise comparisons for the nodes in the URIs. Also, the detection time is considerably lower when the detection does not require exhaustive comparisons or only requires syntactic checking, \textit{e.g.}, \textit{Amorphous URI}, \textit{CRUDy URI}, \textit{Universioned URI}, \textit{Non-standard URI}, and so on. For example, the detection times for \textit{Amorphous URI} antipattern are between 0.001 and 0.089 seconds, the detection times for \textit{Pluralised Nodes} antipattern are between 0.002 and 0.021 seconds. However, the detection times for \textit{Contextless Resource Names} antipattern are between 0.459 (for CISCO IPICS with only five tested URIs) and 351.087 seconds (for Microsoft Azure with 210 tested URIs). And, in this case, the detection times depend on the exhaustive semantic comparisons among the nodes in the URIs.

As shown in Table \ref{Table: Detection Times}, the global average detection time of each antipattern for all IoT APIs is 8.396 seconds. Our detection algorithms for linguistic antipatterns have a polynomial complexity of $\mathcal{O}(n^k)$. The average detection time can be expressed as a polynomial function as shown in Figure \ref{fig:URIwiseAvgDetectTime}.

\begin{center}
\begin{tcolorbox}[colback=gray!15,
                  colframe=gray,
                  width=11cm,
                  left=1pt,
                  right=1pt,
                  top=1pt,
                  bottom=1pt,
                  arc=1mm, auto outer arc]
  \textbf{Summary on RQ$_4$ Efficiency:} In concern to efficiency of \textsf{REST-Ling}, we want to achieve an average detection time for each linguistic antipattern and for each IoT API in the order of seconds. Regardless of the number of URIs tested for each API, \textbf{the \textsf{REST-Ling} had an average detection time of 8.396 seconds}. Moreover, the total detection time for nine antipatterns and nine patterns on 19 APIs with 1,102 URIs was 1,435.727 seconds.
\end{tcolorbox}
\end{center}

\subsection{Threats to Validity}

Our findings may not be generalised to all IoT APIs. However, to minimise the threat to the external validity of our results, we performed experiments on a set of 1,102 URIs from 19 IoT APIs. To minimise the threat to the internal validity, we not only used WordNet~\cite{wordnet} for lexical analyses of URIs, but we also relied on a technique based on the LDA topic modeling to properly capture the context and use the DISCO second-order similarity metric to measure the similarity between the nodes in URIs. But, the outcome of the detection may vary depending on the way the detection heuristics of linguistic patterns and antipatterns are applied because the software engineers may have their own understanding and experience on IoT and on the linguistic patterns and antipatterns. Moreover, in the manual validation, only 91 of the 1,102 URIs were validated manually, which may be not representative. However, to minimise the threat, we aimed for a 95\% confidence level and a confidence interval of 10, thus, we ended up validating 91 URIs and nine linguistic antipatterns, \textit{i.e.}, 819 questions.

In this study, the detection of linguistic antipatterns is performed on the IoT APIs that had well-organised and self-contained documentation. Thus, the detection on IoT APIs with no or very minimal documentation could yield different results. However, we minimised the threats to the construct validity by selecting a set of IoT APIs that are well-documented. Moreover, we tried to minimise the threat to the construct validity by defining detection heuristics after a thorough review of definitions of linguistic patterns and antipatterns. Three professionals were involved in the validation process, and we decided the oracle based on the majority (\textit{i.e.}, when two participants agreed out of the three who participated in the manual validation). The average degree of agreement among the professionals for the manual validation was 0.83. Thus, an average agreement of 0.83 for all antipatterns helps to minimise the threat to the construct validity, \textit{i.e.}, our validation results and accuracy are reliable. The \textsf{REST-Ling} tool currently supports the detection of nine linguistic patterns and nine linguistic antipatterns. The tool has a detection accuracy of more than 80\%. The accuracy of the tool is confirmed via manual validation of the detection outcomes.

In fact, we cannot claim the list of linguistic patterns and antipatterns is complete. Therefore, when talking about linguistic quality of IoT APIs, we refer to the set of linguistic patterns and antipatterns in our study. Nevertheless, to the best of our knowledge, our approach and the empirical study performed on REST IoT APIs is the most comprehensive analysis so far.

To minimise the threats to validity, \textit{i.e.}, to increase the reliability and replicability, we have put all the details of this study online\footnote{\url{https://doi.org/10.5281/zenodo.6393404}}.

\section{Related Work}\label{sec:related}

Analysis techniques to detect linguistic patterns and antipatterns have been previously applied to Object-Oriented (OO) systems' source code \textit{e.g}., \cite{Arnaoudova2016,Abebe2009} to assess their linguistic quality (Section \ref{sec:Syntactic and Semantic Analysis of Source Code}). In addition, several Natural Language Processing (NLP) techniques have been also applied to studying the linguistic quality of APIs and their documentation \textit{e.g.}, \cite{Masse2012,Petrillo2016,Petrillo10.1007978-3-319-94959-816,alliance2012guidelines,10.1007/978-3-319-94959-816, 7930199, Hausenblas2011,Brabra10.1007978}, as discussed in Section \ref{sec:Syntactic and Semantic Analysis of APIs}.

\subsection{Syntactic and Semantic Analysis of Source Code}\label{sec:Syntactic and Semantic Analysis of Source Code}

Abebe \textit{et al.}~\cite{Abebe2009} presented a set of lexicon bad smells in OO code and a tool-suite using semantic analysis techniques to detect them. The authors aimed at improving the linguistic quality of OO source code because high quality and self-descriptive source code comments are useful in developing highly maintainable systems. Khamis \textit{et al.} \cite{khamis2010automatic} proposed the \texttt{JavadocMiner} approach for assessing the quality of in-line documentation relying on heuristics both in terms of language quality and consistency between source code and comments.

Semantic analyses have also been applied to Web services design analysis and development~\cite{Mateos2014,Rodriguez2010}. Rodriguez \textit{et al.}~\cite{Rodriguez2010} presented a study on poor linguistic practices identified on a set of WSDL (Web Service Definition Language) descriptions and provided a catalog of Web services discoverability antipatterns. These antipatterns focus on the comments, elements names, or types used for representing the data models in WSDL documents. Also, Mateos \textit{et al.}~\cite{Mateos2014} presented a tool to detect a subset of the antipatterns proposed in \cite{Rodriguez2010}.

Other researchers also used semantic analyses in different phases of the software development life-cycle~\cite{Arnaoudova2014,Lu2015,Rahman2015}. 
For example, Lu \textit{et al.}~\cite{Lu2015} proposed an approach to improve code searches by identifying relevant synonyms using the WordNet English lexical database \cite{wordnet}. Arnaoudova \textit{et al.}~\cite{Arnaoudova2014} performed a study on identifiers renaming in OO systems. Finally, Rahman and Roy~\cite{Rahman2015} presented an approach to automatically suggest relevant search terms based on the textual descriptions of change tasks in software. These approaches are tailored to OO identifiers and their consistencies with comments~\cite{Arnaoudova2016, Abebe2009} or to traditional SOAP-based Web services interfaces~\cite{Mateos2014,Rodriguez2010}. Therefore, they cannot be applied to IoT APIs due to the peculiarities of their development life-cycle and their consumption nature.

Arnaoudova \textit{et al.}~\cite{Arnaoudova2016} presented the definition of \textit{linguistic antipatterns} and defined 17 linguistic antipatterns in OO programming (\textit{i.e.}, recurring poor practices related to inconsistencies among the naming, documentation, and implementation of a software entity), and implemented their detection algorithms. They searched for the differences between the identifiers used for software entities (\textit{e.g.}, method names and return types) and their implementation and--or documentation. For example, one antipattern is called \textit{``Is'' returns more than a Boolean}, which analyses the name of a method starting with ``Is'' and checks if the method returns a boolean~\cite{Arnaoudova2016}.

Machine learning techniques are also applied in predicting poor design,\textit{ i.e.}, antipatterns. For example, Fakhoury \textit{et al.} \cite{8330265} performed a comparative study to explore how conventional machine learning classifiers perform compared to the deep learning methods in predicting linguistic antipatterns in object-oriented (OO) source code. Aghajani \textit{et al.} \cite{8529834} conducted a large-scale study on more than 1.5k releases of 75 Maven libraries, 14k open-source Java projects based on those libraries, and more than 4k questions related to the libraries from Stack Overflow. More precisely, they studied if client developers are prone to introducing bugs when using APIs involved in linguistic antipatterns. Based on their statistical analysis, it is likely that linguistic antipatterns have an effect on introducing bugs (thus, triggering questions on Stack Overflow) with a probability of 29\%. However, both these studies \cite{8330265, 8529834} were conducted for linguistic antipatterns in OO source code.

\subsection{Syntactic and Semantic Analysis of APIs and their Documentation}\label{sec:Syntactic and Semantic Analysis of APIs}

APIs are the \textit{de facto} standard used by software companies to design, develop, and offer their services through the Internet. Client developers must follow well-documented APIs to use the services and resources offered by those APIs properly. However, there are only a few standards and guidelines that guide API design and development \cite{Masse2012,alliance2012guidelines}. 

Bertolino \textit{et al.} \cite{bertolino2009automatic} modeled the SOAP Web service behavior protocol, \textit{i.e.}, how clients should behave while interacting with the service. They proposed the StrawBerry \cite{bertolino2009automatic} method to automatically derive the Web service behavior protocol from its WSDL interface.

As one of the first studies on APIs, Parrish~\cite{Parrish2010} did a subjective lexical comparison between two well-known APIs, \textit{e.g.}, Facebook and Twitter. The author analysed, for example, the use of verbs and nouns in URIs naming and concluded that developers should rely on nouns instead of verbs while designing REST URIs. 

Masse~\cite{Masse2012} proposed an extensive list of REST API design principles, including the design of URIs, appropriate use of HTTP methods and status codes, meta-data design, and best practices for resource representations. The Open Mobile Alliance (OMA) provides guidelines for designing APIs exhibiting RESTfulness and for properly documenting the APIs, for example, not using verbs as resource identifiers or specifying API version within URIs. Several studies have also been performed in the domain of APIs automatically analyse their structure. For example, Haupt \textit{et al.} \cite{7930199} presented a framework for structural analysis of APIs based on their documentations. They focused on the structural properties of APIs, and later, extended the study towards API governance \cite{Haupt2018}. 

Panziera and Paoli \cite{Panziera:2013:FSR:2487788.2488183} put forward a set of best practices for building self-descriptive REST services, which can be both human-readable and machine-processable (\textit{e.g.}, by using a common vocabulary for REST resources). They proposed a framework to collect information on documentation for  generating descriptions of REST services. They evaluated their framework and reported the accuracy of identifying resources correctly with precision and recall of 72\% and 77\%, respectively. 

Treude \textit{et al.} \cite{treude2015extracting} developed a search-based approach for automatically extracting tasks (\textit{i.e.}, a set of specific programming actions to be undertaken) from software documentation. They tried to minimise the gap between the information needs of the developers' and the documentation structure/content and, thus, assist developers in documentation navigation. Using the suggested approach, which utilises natural language processing techniques, they extracted more than 70\% tasks from two large corpus of software documentation. 

Some studies investigate and analyse services interfaces to measure their linguistic quality, in particular for SOAP Web services \cite{wei2015deriving,bertolino2009automatic} and for APIs \cite{Petrillo2016,Rodriguez2016}. For example, Wei \textit{et al.} \cite{wei2015deriving} presented a framework and algorithms to analyse service interfaces, the SOAP Web services, in particular. They targeted large and overloaded services with the goal to ease their integration and interoperability. The framework enabled to refactor large interfaces and was validated with real commercial logistic systems like FedEx. 

Petrillo \textit{et al.} \cite{Petrillo2016} provided a survey on REST literature and gathered 73 best practices in designing APIs to increase their understandability and reusability. They evaluated three well-known APIs from three Cloud providers, \textit{i.e.}, Google Cloud Platform, OpenStack, and Open Cloud Computing Interface (OCCI), to evaluate their quality based on the identified best practices. 

Rodr{\'i}guez \textit{et al.} \cite{Rodriguez2016} analysed high-volume of REST HTTP traffic, \textit{i.e.}, HTTP requests, to evaluate how well or bad developers implement APIs in practice. They compared the wellness with theoretical Web engineering principles and guidelines. The authors relied on heuristics and metrics to measure the implementation quality by means of antipatterns. Results showed a gap between theory and practice. 

In our previous work \cite{Palma2017SemanticAO}, we proposed the SARA approach for automatically assessing the quality of APIs for Web applications through the detection of linguistic patterns and antipatterns. For the detection of linguistic patterns and antipatterns SARA relied on syntactic and semantic analysis of APIs. In another work \cite{10.1007/978-3-319-94959-816}, we proposed CloudLex and studied the presence of linguistic patterns and antipatterns in 16 cloud computing APIs. The Cloud APIs tend to use heterogeneous terms in their URI designs, and more than half of the URIs were not well-documented. CloudLex showed an average precision of 85\% and a recall of 64\%. In previous works, we also performed studies \cite{Palma2015AreRA, Palma2017SemanticAO} that focused on the `RESTful' aspect of Web APIs, for example, to see if the APIs follow basic REST design principles including (i) statelessness, (ii) cacheability, and (iii) interface uniformity.

In similar lines of research, working with OCCI patterns and antipatterns, Brabra \textit{et al.} \cite{Brabra10.1007978, BRABRA201965} defined a set of patterns and antipatterns, inspired by the OCCI guidelines\footnote{http://occi-wg.org/about/specification/}. They performed an automatic detection of 28 OCCI REST patterns and antipatterns in Cloud APIs by invoking more than 300 operations. 

\subsection{State-of-the-Art Summary}

The analysis of the aforementioned studies allow us to identify some limitations. More specifically, studies dedicated to the OO systems~\cite{Arnaoudova2016,Abebe2009} or SOAP-based Web services interfaces~\cite{Mateos2014,Rodriguez2010} are not applicable to APIs for IoT applications. Although, there are guidelines on API design~\cite{Masse2012}, the semantic aspects of API design were considered in very few works \cite{7930199,Haupt2018}. Some studies analysed the APIs or their documentation but did not assess the linguistic quality of the APIs (e.g., \cite{Petrillo2016,bertolino2009automatic,wei2015deriving,Rodriguez2016}) or software documentation (e.g., \cite{treude2015extracting}). Other works only focused on the structural design of the APIs, e.g., \cite{Brabra10.1007978, BRABRA201965}.

Although some of the aforementioned approaches dealt with linguistic aspects of REST or cloud computing APIs, in most cases, they only relied on the subjective view of a set of good linguistic practices and recommendations. There is a lack of dedicated approaches that automatically assess the linguistic quality of APIs from IoT providers by detecting both poor and best practices. Other research focused on the analysis of linguistic aspects of the APIs and their documentation, and to the best of our knowledge, our study is the first that focuses on the linguistic design quality of APIs from IoT providers.

\begin{table}[t!]
\caption{Comparison with the relevant studies in the literature.}\label{tab:comparison}
\def\arraystretch{1.4}
\setlength\tabcolsep{2pt}
\centering
\scriptsize
\begin{tabular}{l p{2.5cm} p{2.5cm} p{4cm}}
\toprule
\multicolumn{1}{c}{\multirow{1}{*}{\textbf{Study}}} & \multicolumn{1}{c}{\multirow{1}{*}{\textbf{Goal of the Study}}} & \multicolumn{1}{c}{\multirow{1}{*}{\textbf{Target APIs}}}  & \multicolumn{1}{c}{\multirow{1}{*}{\textbf{Analysis Method}}} \\  \midrule
Parrish \cite{Parrish2010} & Lexical analysis of 2 social network APIs & Facebook and Twitter & Identification of verbs and nouns  \\ \midrule
Wei \textit{et al.} \cite{wei2015deriving} & Structural analysis of the service interfaces & 272 operations from 13 cloud services related to Amazon, FedEx, etc. & Number of operations each service provides, average number (per operation) of input parameters, output parameters, business entities, etc. \\ \midrule
Brabra \textit{et al.} \cite{Brabra10.1007978} & Definition and detection of OCCI REST patterns and antipatterns & 6 APIs for Cloud services & Detection of 28 OCCI REST antipatterns and their corresponding patterns  \\ \midrule
Petrillo \textit{et al.} \cite{Petrillo2016} & Check APIs for conformance to 73 best practices & 3 Cloud APIs (Google Cloud Platform, OpenStack, and OCCI) & For each Cloud API, manually checking the conformance to best practices at the service level  \\ \midrule
Rodriguez \textit{et al.} \cite{Rodriguez2016} & Check APIs for compliance or violation to 6 standardised practices & 78 GB of HTTP traffic from Telecom Italia & For each action (Post, Get, Put, Delete, etc.) check for the conformance with standardised semantics, and also structural design of request URIs  \\ \midrule
Haupt \textit{et al.} \cite{7930199} & Structural analysis of APIs & 286 Swagger API description documents & Size of APIs in terms of resources, Number of POST/DELETE methods, read only resources, etc. \\ \midrule
Palma \textit{et al.} \cite{Palma2017SemanticAO} & Detection of linguistic antipatterns in APIs & 18 APIs for Web applications & Heuristics-based detection for 6 linguistic antipatterns and their corresponding patterns \\ \midrule
Petrillo \textit{et al.} \cite{Petrillo10.1007978-3-319-94959-816} & Linguistic quality assessment of Cloud Computing APIs & 23,062 URIs from the 16 Cloud API providers & Detection of 2 linguistic antipatterns and their corresponding patterns \\ \midrule
Brabra \textit{et al.} \cite{BRABRA201965} & Detection of OCCI REST patterns and antipatterns & 5 Cloud APIs including OCCI, COAPS, OpenNebula, Amazon S3, and Rackspace & Detection of 21 OCCI REST antipatterns and their corresponding patterns  \\ \midrule
SARAv2 & Detection of linguistic antipatterns and patterns in APIs from IoT providers & 19 APIs from 18 IoT providers & Heuristics-based detection for 9 linguistic antipatterns and their corresponding patterns  \\ \bottomrule
\end{tabular}
\end{table}


Table \ref{tab:comparison} shows a summary of the comparison between our SARAv2 approach and the related state-of-the-art studies in terms of their goals and methods. In providing a big picture of the comparison: firstly, SARAv2 is a general approach for analysing REST APIs, and the empirical experiment in the current paper is the first study related to IoT APIs. In this aspect, we studied 19 APIs from 18 different IoT providers, where we performed both syntactic and semantic analysis of more than 1,100 URIs. We performed the detection of nine linguistic antipatterns and their corresponding nine linguistic patterns to assess the linguistic quality of IoT APIs because we conjecture that poor linguistic quality hinders the consumption, reusability, and maintenance and evolution of APIs. Studies have been performed for Cloud services (\textit{e.g.}, \cite{Petrillo2016, Petrillo10.1007978-3-319-94959-816, Brabra10.1007978}) or REST Web services (\textit{e.g.}, \cite{Palma2015AreRA, Parrish2010, Rodriguez2016}), which are mostly based on syntactic analysis. However, to the best of our knowledge, SARAv2 is the first study that analyse IoT APIs both syntactically and semantically.

Secondly, our analysis involved 19 APIs from 18 different IoT providers, which is also comparatively higher than any other studies in the literature, \textit{i.e.}, we wanted to investigate a set of APIs from heterogeneous providers to see, on an average, the ratio of well-designed and poorly-designed APIs in terms of linguistic quality. 

\begin{table}[ht!]
    \centering
    \caption{Comparison of the detection accuracy with the state-of-the-art approaches.}
    \label{tab:accuracyComparisonRQ3}
    \scriptsize
    \setlength{\tabcolsep}{3pt}
    \begin{tabular}{lcccccccccc}
    \toprule
Study & \rotatebox{90}{Tidy/Amorphous URIs} & \rotatebox{90}{Contextualised/Contextless Resource} & \rotatebox{90}{Verbless/CRUDy URIs} & \rotatebox{90}{Hierarchical/Non-hierarchical Nodes} & \rotatebox{90}{Singularised/Pluralised Nodes} & \rotatebox{90}{Pertinent/Non-pertinent Documentation} & \rotatebox{90}{Consistent/Inconsistent Documentation} & \rotatebox{90}{Unversioned/Versioned URI} & \rotatebox{90}{Non-standard/Standard URI} & \rotatebox{90}{Overall Accuracy/Precision}  \\
\midrule
Parrish \cite{Parrish2010} & - & - & - & - & - & - & - & - & - & \emph{n/a}  \\
Wei \textit{et al.} \cite{wei2015deriving} & - & - & - & - & - & - & - & - & - & 91\%  \\
Brabra \textit{et al.} \cite{Brabra10.1007978} & - & - & 100\% & - & - & - & - & - & 100\% & 98\%  \\
Petrillo \textit{et al.} \cite{Petrillo2016} & - & - & - & - & - & - & - & - & - & \emph{n/a}  \\
Rodriguez \textit{et al.} \cite{Rodriguez2016} & - & - & - & - & - & - & - & - & - & \emph{n/a}  \\
Haupt \textit{et al.} \cite{7930199} & - & - & - & - & - & - & - & - & - & \emph{n/a} \\
Palma \textit{et al.} \cite{Palma2015AreRA} & 96\% & 77\% & 95\% & 60\% & 73\% & - & - & - & - & 81\% \\
Palma \textit{et al.} \cite{Palma2017SemanticAO} & 98\% & 90\% & 100\% & 60\% & 73\% & 66\% & - & - & - & 84\%  \\
Petrillo \textit{et al.} \cite{Petrillo10.1007978-3-319-94959-816} & - & 71\% & - & - & - & 77\% & - & - & - & 74\%  \\
Brabra \textit{et al.} \cite{BRABRA201965} & 100\% & - & 100\% & - & 100\% & - & - & - & - & 100\%  \\
This Study (SARAv2) & 67\% & 78\% & 97\% & 85\% & 88\% & 41\% & 84\% & 100\% & 89\% & 81\% \\
\bottomrule
\multicolumn{10}{l}{\textit{*Values that are not reported by the studies denoted by `-'.}}\\
\multicolumn{10}{l}{\textit{*The \textit{n/a} refers to the cases where the precision/accuracy are not reported.}}
    \end{tabular}
\end{table}

A final comparison can be made from the perspective of detection accuracy. Our SARAv2 approach performs with an average accuracy of more than 80\%. Nevertheless, other studies (as reported in Table \ref{tab:accuracyComparisonRQ3}) show an average precision between 80.9\% and 100\%. However, these studies are (i) either focus on other types of APIs (\textit{i.e.}, Cloud services or Web APIs) or (ii) the number of analysed APIs is low (\textit{i.e.}, between 2 and 15 APIs) or (iii) the number of detected linguistic patterns and antipatterns is relatively lower (\textit{i.e.}, between 2 and 28 antipatterns) or (iv) they only perform very fine-grained syntactic analyses (\textit{i.e.}, those for OCCI patterns and antipatterns \cite{Petrillo2016, Brabra10.1007978}). Considering the highly semantic nature of our automatic analysis using the SARAv2 approach and the subjective validation of the results using experts, which might significantly differ given the full degree of freedom for deciding patterns and antipatterns, we consider an average accuracy of more than 80\% is acceptable in the domain of natural language processing \cite{Shah:2015:RAN:2815021.2815032}.
\section{Conclusion and Future Work}\label{Conclusion and Future Work}

The understandability and reusability are two critical factors for API providers. In the literature, researchers analysed APIs for Web applications and cloud services to assess their linguistic design quality ~\cite{Petrillo2016, Arnaoudova2016, Petrillo10.1007978-3-319-94959-816, 10.1007/978-3-319-94959-816, 7930199, Hausenblas2011, Brabra10.1007978}. In this study, we assess the linguistic quality of APIs for IoT applications by analysing whether they contain linguistic antipatterns. We proposed the SARAv2 (Semantic Analysis of REST APIs version two) approach and used it to perform syntactic and semantic analyses of REST APIs for IoT applications. The \textsf{REST-Ling} realises the SARAv2 approach as a web application to automate the detection of linguistic patterns and antipatterns.

We utilised the \textsf{REST-Ling} tool to detect nine linguistic patterns and antipatterns. We validated the \textsf{REST-Ling} tool by analysing 1,102 URIs from 19 REST APIs for IoT applications and showed its accuracy of over 80\%. 

From the 19 analysed APIs, we found that all of them organise URIs nodes in a hierarchical manner and only Caret, CubeSensors, and Droplit.io APIs involve syntactical URIs design problems. Moreover, IoT APIs designers, in general, do not use CRUDy terms in URIs, which is a good design practice, but then again, they tend not to use versioning in URI -- a poor practice. Also, most designers in IoT use hierarchical organisation of nodes in URIs and document them using consistent language. Further, the \textcolor{red}{\begin{scriptsize}\ding{54}\end{scriptsize}}\textit{Non-pertinent Documentation} was common in all IoT APIs and the majority of the APIs had \textcolor{red}{\begin{scriptsize}\ding{54}\end{scriptsize}}\textit{Unversioned URI}. In contrast, most of the APIs followed \textcolor{teal}{\begin{scriptsize}\ding{52}\end{scriptsize}}\textit{Tidy URI} and \textcolor{teal}{\begin{scriptsize}\ding{52}\end{scriptsize}}\textit{Consistent Documentation}.

As we compare the detection of antipatterns across the domains, we found that the APIs for Web applications are highly prone to \textcolor{red}{\begin{scriptsize}\ding{54}\end{scriptsize}}\textit{Amorphous URI} although carefully implement patterns like \textcolor{teal}{\begin{scriptsize}\ding{52}\end{scriptsize}}\textit{Verbless URI} and \textcolor{teal}{\begin{scriptsize}\ding{52}\end{scriptsize}}\textit{Pertinent Documentation}. We also found that the IoT APIs have very tidy URIs, \textit{i.e.}, follow \textcolor{teal}{\begin{scriptsize}\ding{52}\end{scriptsize}}\textit{Tidy URI} and the nodes in the URIs are organised hierarchically, \textit{i.e.}, follow \textcolor{teal}{\begin{scriptsize}\ding{52}\end{scriptsize}}\textit{Hierarchical Nodes}. On an average, 34\% of the URIs from the APIs for Web applications are detected as having linguistic antipatterns, in contrast, 17\% of the URIs from the IoT APIs are detected as antipatterns. As for the linguistic patterns, 73\% URIs in IoT APIs are well-designed compared to 42\% URIs of APIs for Web applications, which suggests that IoT APIs like Amazon AWS, Google Nest, IBM Watson, Microsoft Azure are comparatively more well-designed than the APIs for general-purpose Web applications like Facebook, YouTube, or Instagram.

As future work, we want to apply the SARAv2 approach, thus, the REST-Ling tool, to other IoT APIs. Recently, OpenAPI has been evolved to the industry standard for REST API design and specification. We want to analyse OpenAPI JSON/YAML specifications to assess their design and documentation quality. We also want to investigate two of the patterns and antipatterns further -- \textit{Pluralised} \textit{vs.} \textit{Singularised Nodes} and \textit{Non-pertinent} \textit{vs.} \textit{Pertinent Documentation} -- as they are affected more by the cognitive ability of the client developers. We also want to build and include an IoT-specific ontology to perform an improved semantic analysis. Finally, while comparing the detection of SARA and SARAv2, a further extension could be to compare the services from the same company/team (\textit{e.g.}, the REST APIs for Web applications \textit{vs.} IoT APIs for IoT applications from Microsoft or Google) to see whether the difference in the antipatterns is due to the difference of the domain (IoT \textit{vs.} Web) or due to different companies having different API design principles or level of experience.
\section*{Acknowledgments}
We would like to thank Niklas Emevi, a full stack Web developer at Tieto CEM, for taking part in the validation process. We are thankful to Osama Zarraa and Ahmad Sadia for their contributions in developing the tool. We extend gratitude to The Knowledge Foundation that partially supported this research through the SHADE H\"{O}G-project 2017/0176. This study was also conducted with the support from Linnaeus University Centre for Data Intensive Sciences and Applications (DISA).

\begin{small}
\bibliographystyle{elsarticle-num} 
\bibliography{JSS}
\end{small}

\end{document}